\documentclass[%
 aip,
 amsmath,amssymb,
 reprint,%
]{revtex4-1}

\usepackage{graphicx}
\usepackage{dcolumn}
\usepackage{bm}
\usepackage{breqn}
\makeatletter
\let\cat@comma@active\@empty
\makeatother

\usepackage[utf8]{inputenc}
\usepackage[T1]{fontenc}
\usepackage{mathptmx}
\usepackage{etoolbox}

\usepackage{xcolor}
\usepackage{soul}
\usepackage[colorlinks=true,linkcolor=blue,citecolor=blue,urlcolor=blue]{hyperref}

\newcommand*{\figreff}[2][]{%
  \hyperref[{fig:#2}]{%
    Figure~\ref*{fig:#2}%
    \ifx\\#1\\%
    \else
      \,#1%
    \fi
  }%
}

\newcommand*{\figref}[2][]{%
  \hyperref[{fig:#2}]{%
    Fig.~\ref*{fig:#2}%
    \ifx\\#1\\%
    \else
      \,#1%
    \fi
  }%
}

\newcommand*{\tabref}[2][]{%
  \hyperref[{tab:#2}]{%
    Table.~\ref*{tab:#2}%
    \ifx\\#1\\%
    \else
      \,#1%
    \fi
  }%
}

\makeatletter
    \def\@email#1#2{%
 \endgroup
 \patchcmd{\titleblock@produce}
  {\frontmatter@RRAPformat}
  {\frontmatter@RRAPformat{\produce@RRAP{*#1\href{mailto:#2}{#2}}}\frontmatter@RRAPformat}
  {}{}
}%
\makeatother
\begin{document}

\preprint{AIP/123-QED}

\title{The Diffuse Solid Method for Wetting and Multiphase Fluid Simulations in Complex Geometries}
\author{Fandi Oktasendra}
 \affiliation{Department of Physics, Durham University, Durham DH1 3LE, United Kingdom}
\affiliation{Department of Physics, Universitas Negeri Padang, Padang 25131, Indonesia }

\author{Michael Rennick}%
\affiliation{Department of Physics, Durham University, Durham DH1 3LE, United Kingdom}
\affiliation{Institute for Multiscale Thermofluids, School of Engineering, University of Edinburgh, Edinburgh EH9 3FD, United Kingdom}

\author{Samuel J. Avis}%
\affiliation{Department of Physics, Durham University, Durham DH1 3LE, United Kingdom}

\author{Jack R. Panter}%
\affiliation{School of Engineering, Mathematics and Physics, University of East Anglia, Norwich NR4 7TJ, United Kingdom}
 \email{jack.panter@uea.ac.uk}

\author{Halim Kusumaatmaja}
\affiliation{Department of Physics, Durham University, Durham DH1 3LE, United Kingdom}
\affiliation{Institute for Multiscale Thermofluids, School of Engineering, University of Edinburgh, Edinburgh EH9 3FD, United Kingdom}
\email{halim.kusumaatmaja@ed.ac.uk}

\date{\today}

\begin{abstract}
We develop a diffuse solid method that is versatile and accurate for modeling wetting and multiphase flows in highly complex geometries. In this scheme, we harness $N+1$-component phase field models to investigate interface shapes and flow dynamics of $N$ fluid components, and we optimize how to constrain the evolution of the component employed as the solid phase to conform to any pre-defined geometry. Implementations for phase field energy minimization and lattice Boltzmann method are presented. Our approach does not need special treatment for the fluid-solid wetting boundary condition, which makes it simple to implement. To demonstrate its broad applicability, we employ the diffuse solid method to explore wide-ranging examples, including droplet contact angle on a flat surface, particle adsorption on a fluid-fluid interface, critical pressure on micropillars and on \textit{Salvinia} leaf structures, capillary rise against gravity, Lucas-Washburn’s law for capillary filling, and droplet motion on a sinusoidally undulated surface. Our proposed approach can be beneficial to computationally study multiphase fluid interactions with textured solid surfaces that are ubiquitous in nature and engineering applications. 
\end{abstract}

\maketitle

%

\section{\label{sec:level1}Introduction \protect} 

Inspired by the extraordinary diversity of surface structures adapted to manipulating liquids in nature~\cite{Barthlott2017,Comanns2018}, there is a rapid and increasing drive to fabricate complex physical and chemical textures with functional wetting properties~\cite{Arzt2021}. This is particularly prevalent in the field of surface science, which has seen a range of recent innovations across fog harvesting~\cite{Yu2022}, robust liquid-repellent surfaces~\cite{Chen2022}, directional wetting~\cite{Hou2023}, and wettability control for sensing~\cite{Vineeth2023}. This has been enabled by advances in fabrication methods of increasingly complex surface microstructures, such as additive manufacturing (3D printing)~\cite{Yan2020}, photolithography~\cite{Gao20155918} and self assembly~\cite{Brassat2020}. However, the interest in understanding wetting interactions with complex surface structures is not only limited to fundamental wetting physics, but spans a diverse range of fields, from open microfluidics in biomedicine ~\cite{Berthier2019}, to sub-milimeter manufacturing in engineering ~\cite{Kwok2020}, and carbon storage in geology ~\cite{Liu2024}.

To complement experimental and theoretical efforts, there is a need for efficient and accurate simulation methods that are capable of modeling the statics and dynamics of capillary phenomena on highly complex solid structures. Ideal simulation methods must achieve the following aims: (1) accuracy in modeling surface tensions and contact angles, (2) generality and accuracy in describing the solid geometry, and (3) computational efficiency. In this work, to achieve these aims, we present the Diffuse Solid Method which is applied both to the Lattice Boltzmann Method (for fluid dynamics), and to a phase-field energy minimization method (for static equilibria). In this, we use a $N+1$-phase diffuse interface method that models $N$ evolving fluid components and 1 solid component that we limit the evolution of. A principal outcome is to show how to control this partial evolution of the solid to conform to any user-defined geometry, without compromising the accuracy of the liquid contact angles. While we demonstrate the scheme for $N=2$, the scheme can be easily generalized to an arbitrary number of components. Importantly, the interfacial energy balance is intrinsically accounted for by the phase field free energy, and no separate scheme is required for the fluid-solid wetting boundary condition.

Previously, a number of alternative simulation methods have been developed towards achieving accurate wetting simulations on curved and complex boundaries.
At the highest level, these simulation strategies are split into two classes: sharp interface models, where the fluid interface is modeled as infinitely thin and explicitly tracked during simulations~\cite{Chamakos20211647,yue2010sharp}; and diffuse interface models, where each fluid phase is associated with a smoothly varying order parameter, and fluid-fluid interfaces have a characteristic thickness~\cite{Jacqmin1999,Anderson1998139}. Sharp interface models are often more efficient computationally, however, they are cumbersome for problems that entail topological changes in the fluid interface shapes, such as due to droplet break-up and coalescence, as well as interface pinning and depinning events~\cite{Anderson1998139,Lee2002492,Lee2002514,badalassi2003computation,li2016pinning}. Since these effects are essential to capture for the problems of interest in this work, here we shall focus on diffuse interface models.

For diffuse interface models, there are already well-established methods for incorporating fluid-solid interactions on flat surfaces~\cite{Ledesma-Aguilar20148267,Raman2016336,Dong2020,gong2015numerical}, including liquid contact angle and contact line pinning and depinning due to regions of different surface chemistry. However, current extensions to complex, and often curved, solid shapes are either cumbersome or introduce unwanted artifacts. This is a principal issue when employing diffuse interface models on methods that are based on fixed, regular grids, such as the Lattice Boltzmann Method. However, it is desirable to use such methods as
they are highly efficient for parallel and GPU computing, and hence are becoming increasingly popular for solving large-scale simulation challenges~\cite{Shan19931815,Ye2015114,Tolke2002535}.

Within the Lattice Boltzmann Method, a common approach to multiphase flows on curved and complex solids is to treat the surface via the conceptually simple staircase approximation,~\cite{Stahl20101625,Lu2017268,Yu2020}, or similarly, as in the case of the Homogenised Lattice Boltzmann Method, by defining voxellated regions of space within the solid body to be momentum sinks~\cite{Lautenschlaeger2022}. This, however, leads to interface pinning at the corners of the approximated shapes, and in problems where surface energies are important (such as capillary rise as we shall demonstrate), the solid-fluid interaction area is dramatically overestimated. 

A key strategy to enable simulations on solid surfaces which do not conform to the underlying (Eulerian) mesh, is the Immersed Boundary Method, in which the solid surface is described via a superimposed non-conformal (Lagrangian) mesh. Recent advances have overcome core challenges to this method for use in multiphase flows (namely poorly obeyed mass conservation and limited stable fluid density ratios) that now allow accurate wetting boundary conditions~\cite{Yao2022,Li2024}. Nonetheless, an inherent downside of the Immersed Boundary Method is the need for 
additional interpolation steps between the Eulerian and Lagrangian meshes. 

A contrasting approach which entirely avoids the explicit description of the solid boundary is Fluid Particle Dynamics, in which the solid is modeled as a highly viscous liquid~\cite{Tanaka20001338,Furukawa20183738}. A core advantage of this method is that no special wetting boundary conditions are required: the correct contact angles emerge implicitly from the diffuse interface interactions. Although this method is successful for, e.g. colloidal systems, where the solids are always spherical, this cannot achieve solid geometries with arbitrary user-defined shapes.

Sitting between the solid and liquid methods is the Diffuse Domain approach, currently employed in Cahn-Hilliard and Navier-Stokes solvers~\cite{Aland2010}. In this, the solid domain is identified with an order parameter, exactly as for the fluid phases. In contrast to Fluid Particle Dynamics however, the solid order parameter is not updated in time, and is instead fixed upon initialization. In order to accurately capture the contact angles on the solid phase, it is essential to have the correct solid-fluid diffuse interfacial profiles. Recently, this has been achieved through using a pre-simulation stage to relax the solid order parameter via a diffusion process ~\cite{YANG2023112345,ZHU2025113699}. However, the Diffuse Domain approach suffers from an inherent limitation: relaxing the solid diffuse interface also allows the overall shape of the solid object to evolve away from the desired geometry. A principal achievement in our work is to develop a method of constraining the solid morphology, while allowing the correct diffuse interface to form. This method is essential in order to accurately model wetting on complex geometries, and to allow implementation in free energy minimization and the Lattice Boltzmann methods.

Following the description of our Diffuse Solid Method in Sec.~\ref{sec.2}, we rigorously demonstrate it's broad applicability and accuracy across a range of benchmarks. In Sec.~\ref{sec.3}, static phenomena are investigated, featuring droplets on curved surfaces as a test of the contact angle accuracy, equilibrium capillary rise to test the energy contribution accuracy, and critical intrusion pressures on micropillared surfaces as a test of the pressure implementation. A case study is then shown, to show how our method can be applied to understand the wetting properties of nature-inspired surface structures. In Sec.~\ref{sec.4}, we advance to study dynamic phenomena via the Lattice Boltzmann Method: capillary filling in micro-channels, and droplet motion on an undulating surface.

\section{\label{sec.2}Diffuse Solid Method}

\subsection{Phase Field Energy Minimization}

The diffuse solid method uses a phase field model with $N+1$ immiscible components in which the local composition of each component is represented by an order parameter $C_n$, where $n = \{ 1,\cdots, N+1 \}$. Here we focus on $N=2$ for concreteness.  In this context, we designate one component, $C_1$, as a solid phase and the rest, $C_2$ and $C_3$, as fluids which shall be referred to as liquid and gas phases, respectively.

The ternary model ($N+1=3$) used in this work is based on the work of Semprebon, \textit{et al.}\cite{Semprebon2016}, in which the thermodynamics of the system are described by a free-energy functional of the fluids that captures the immiscibility of the fluid components and the surface tensions between different fluids. The phase interface is modelled using a diffuse interface with a finite thickness, which yields a smooth transition region between two different phases.

For the phase field energy minimization approach, the total free energy of the system can be written as 
\begin{equation}
\Psi(C_n) = \int_V (\Psi_{\textrm{b}} + \Psi_{\textrm{i}} + \Psi_{\textrm{cs}} + \Psi_{\textrm{cf}}) \textrm{d}V,
\label{frozen:total_energy}
\end{equation}
where all of the energy densities are integrated over the system's volume, $V$. The bulk energy density of the phases, $\Psi_{\textrm{b}}$, is modeled using a double well potential
\begin{equation}
\Psi_{\textrm{b}} = \sum_{n=1}^{3}\frac{\kappa_n}{2}C_n^2(1-C_n)^2.
\label{frozen:bulk_energy}
\end{equation}
Here, the $\kappa_n$'s are tunable parameters related to the interfacial tension between different fluids. The form of potentials chosen in this model has two minima at $C_n = 0$ and $1$. We then relate $C_n = 0$ to indicate the absence of component $n$ and $C_n = 1$ to indicate pure component $n$. We also normalise the total composition to unity so that $C_1+C_2+C_3 = 1$ to obtain $C_3= 1 - C_1- C_2$, which reduces the number of the order parameter in the system that needs to be optimised. It is worth noting that this constraint allows $C_1$ and $C_2$, in principle, to have any value. 

$\Psi_{\textrm{i}}$ is referred to as the gradient energy density, which accounts for the energy penalty for having interfaces. The gradient term takes the form
\begin{equation}
\Psi_{\textrm{i}} = \sum_{n=1}^{3}\frac{\kappa_n'}{2}|\boldsymbol{\nabla} C_n|^2,
\label{frozen:interface_energy}
\end{equation}
where $\kappa_n'$'s are also tunable parameters whose values determine the contribution of the gradient terms to the total free energy. 

In combination, $\Psi_{\textrm{b}}$ and $\Psi_{\textrm{i}}$ energetically penalize the formation of interfaces between each pair of phases, such that the surface tension between phases $n$ and $m$, $\gamma_{nm}$, is given by
\begin{equation}
\gamma_{nm} = \frac{\alpha}{6} (\kappa_n + \kappa_m).
\label{frozen:interface_tension}
\end{equation}
$\alpha$ is the characteristic interface width,
\begin{equation}
\alpha=\sqrt{(\kappa_n'+\kappa_m')/(\kappa_n+\kappa_m)},
\end{equation}
that describes the smooth variation of $C_n$ across an interface. For a planar interface (e.g. spanning the $z-y$ plane) between phases $n$ and $m$, this profile takes the form
\begin{equation}
C_n(x) = \frac{1}{2}+\frac{1}{2}\tanh{\frac{x}{2\alpha}}.
\label{frozen:interface_profile}
\end{equation}
Here, the value of $C_n$ varies between 0 and 1 across the interface, and we take the value of $C_n = 0.5$ as the interfacial boundary between the two phases. We can further use the definition of $\alpha$ to simplify the simulation variables by relating $\kappa_{\{n,m\}}'$ and $\kappa_{\{n,m\}}$ via $(\kappa_n'+\kappa_m') = \alpha^2 (\kappa_n+\kappa_m)$. In this work, we take $\kappa_n' = \alpha^2 \kappa_n$ and use $\alpha = 1$ (this and all other parameters are in lattice units, L.U., unless otherwise stated). While we will call $\alpha$ the interface width, it is noted that for $\alpha = 1$, the transition between $C_n = 0.1$ and $C_n = 0.9$ occurs over a range of $4.4\alpha$.

In order to relate the phase field model parameters $\kappa$ to physical interfacial characteristics (namely the surface tensions $\gamma$), Eq.~(\ref{frozen:interface_tension}) can be inverted to yield,
\begin{equation}
\begin{split}
    \kappa_1 &= \frac{3}{\alpha}(\gamma_{12}+\gamma_{13}-\gamma_{23}), \\
    \kappa_2 &= \frac{3}{\alpha}(\gamma_{12}-\gamma_{13}+\gamma_{23}), \\
    \kappa_3 &= \frac{3}{\alpha}(-\gamma_{12}+\gamma_{13}+\gamma_{23}).
    \label{frozen:kappa}
\end{split}
\end{equation}

$\Psi_{\textrm{cs}}$ is the constraining energy density that allows the system to preserve either the pressure difference between liquid and gas phases, $\Delta P$, or the volume of the liquid phase, $V_{\textrm{l}}$, during the minimisation. Following Oktasendra, {\it et al.}~\cite{Oktasendra2023}, when the former is desired, the constraining potential takes the form, 
\begin{equation}
\Psi_{\textrm{cs}} = -\Delta P \, V_{\textrm{l}} ,
\label{frozen:pressure_const}
\end{equation}
whereas for the latter,
\begin{equation}
\Psi_{\textrm{cs}} = \frac{1}{2}k_{\text{cs}}(V_{\textrm{l}} - V_{\textrm{0}})^2.
\label{frozen:volume_const}
\end{equation}
Here, $k_{\text{cs}} > 0$ is a constant, and $V_{\textrm{0}}$ is the target volume. The volume of the liquid phase, $V_{\textrm{l}}$ can be calculated as 
\begin{equation}
V_{\textrm{l}} = \int_V C_2 \textrm{d}V.
\label{frozen:volume_liquid}
\end{equation}

Thus far, the described model is similar to that we have previously published for phase field energy minimisation~\cite{Oktasendra2023,panter2017impact,panter2019multifaceted,sadullah2020factors}. For the diffuse solid method, a key addition is the confining potential $\Psi_{\textrm{cf}}$. This potential acts only on the `solid' phase, which throughout we choose to be phase 1. $\Psi_{\textrm{cf}}$ must achieve two purposes (1) to confine phase 1 to a defined region of arbitrary geometry (which we call the `solid region', $S$), (2) to allow the interface of phase 1 to relax into the equilibrium diffuse interfacial profile. This latter point is essential to accurately define the surface tensions with the solid phase, as defined in Eq.~(\ref{frozen:interface_tension}). These can both be achieved by defining
\begin{equation}
        \Psi_{\textrm{cf}}(\mathbf{r}) =
        \begin{cases}
            \psi_{\textrm{cf}} \quad \mathrm{if} \quad C_1 > 0.5 \quad\mathrm{and} \quad \mathbf{r} \in S, \\
             \psi_{\textrm{cf}} \quad \mathrm{if} \quad C_1 < 0.5 \quad\mathrm{and} \quad \mathbf{r} \notin S, \\
             0 \quad \mathrm{otherwise},
        \end{cases}
\label{frozen:poly-quadratic}
\end{equation}
where $\mathbf{r}$ defines the position of each point in the system, and $\psi_{\textrm{cf}}=\beta(C_1 - 0.5)^2$. $\beta$ is a positive variable controlling the strength of the potential. This form of $\Psi_{\textrm{cf}}$ energetically penalizes the solid phase from moving out of the solid region, and likewise energetically penalizes fluid phases from entering the solid region, so centering the solid-fluid interface at the solid region boundary.

\begin{figure}[h!]
	\centering
	\includegraphics[width=8.6cm]{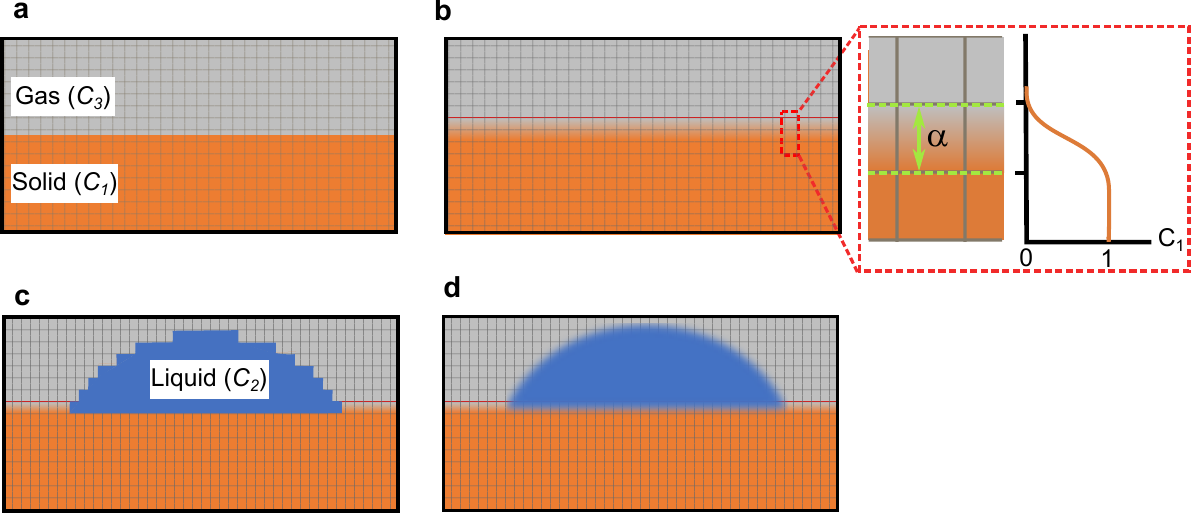}
	\caption{Schematic of implementing the diffuse solid method for a droplet on a flat surface in the energy minimization. (a) Initialization of the flat solid geometry ($C_1$) and the gas phase ($C_3$). (b) The diffuse solid interface with a thickness of $\alpha$ is formed after a number of iterations as highlighted by the inset which also shows the tanh profile of the solid component ($C_1$) across the interface. (c) Initialisation of the liquid phase ($C_2$) using a staircase approximation. (d) The final result after the minimization process is completed.}
	\label{fig:schematic_model}
\end{figure}

In the simulation, we discretize $\Psi$ in Eq.~(\ref{frozen:total_energy}) into $L_x \times L_y \times L_z$ cubic lattice points (called nodes), in which each node is associated with a value of $\{C_n\}_{ijk}$, where the indices $ i,j,k$ indicate the spatial dimension in the \textit{x-}, \textit{y-}, and \textit{z-}directions. The grid spacing, $G$, is set to be equal to the interface width $\alpha$, which is taken to be 1. Depending on the case being investigated, we employ two types of boundary conditions: the periodic and symmetry boundary conditions. The energy minimization is then carried out using the computationally efficient L-BFGS algorithm~\cite{Nocedal1980951,Liu1989503}, optimized for both memory usage and processing speed. During the minimization routine, the value of $\{C_n\}_{ijk}$ of each node is updated in order to achieve the minimum total energy configuration. 

The implementation of the diffuse solid method during the energy minimization process in the simulation is carried out in two stages, as illustrated in ~\figref{schematic_model}. In the first stage, energy minimization is performed to create a solid geometry with a smooth and diffuse solid interface\cite{Panter2023}. The solid is initialized as component 1 ($C_1$) within the desired geometry. The remaining system is filled with pure gas ($C_3$), and the liquid phase ($C_2$) is fixed at $0$ (\figref{schematic_model}(a)). The total free energy is subsequently minimized using the L-BFGS algorithm. During this stage, the energy minimization process is truncated at a specific number of iterations (typically, $n_{\text{iter}} = 10$) to allow the solid-gas diffuse interface to form without strongly perturbing the intended solid geometry's shape (\figref{schematic_model}(b)). Once this stage is complete, the liquid phase ($C_2$) is initialized in the system according to a predefined configuration (\figref{schematic_model}(c)). The solid phase ($C_1$) is now fixed, while the liquid ($C_2$) and gas ($C_3$) phases are allowed to vary. The total free energy is now minimized under these constraints (\figref{schematic_model}(d)). Convergence of the minimization process is determined based on an \textit{rms} gradient condition: convergence is achieved when $|\boldsymbol{\nabla} \Psi| < 10^{-5}$.

The determination of $n_{\text{iter}}$, $\alpha$, $\beta$, and the domain dimensions are chosen to balance the accuracy of the solid geometry, accuracy of the wetting properties described in Section \ref{sec.3}, and computational time. A systematic investigation of the impact of each of these parameters is performed in Supplementary Material, Section  A.

\subsection{Lattice Boltzmann Method}

Next, we will describe how the diffuse solid method can be integrated with lattice Boltzmann method to model wetting dynamics involving solid boundaries of complex shapes. Generally, the continuum equations of motion solved by the lattice Boltzmann method comprise of three equations. First, the continuity equation
\begin{equation}
\partial_t\rho+\mathbf{\boldsymbol{\nabla}}\cdot(\rho \mathbf{v}) = 0,
\label{eqn:ce}
\end{equation}
which describes the conservation of mass in a fluid with density $\rho$ and velocity $\mathbf{v}$. 

Second, the incompressible Navier-Stokes equation
\begin{equation}
\partial_t(\rho \mathbf{v}) + \boldsymbol{\nabla}\cdot(\rho\mathbf{vv})= - \mathbf{\boldsymbol{\nabla}} \cdot P + \mathbf{F} + \eta\Delta \mathbf{v},
\label{eqn:nse}
\end{equation}
which describes the conservation of momentum. Here, $P$ denotes the pressure, $\eta$ is the dynamic viscosity and $\mathbf{F}=\mathbf{F}_s+\mathbf{F}_g$ is the body force applied to the fluid. $\mathbf{F}_s$ is the surface tension force between the two fluid components in the system and $\mathbf{F}_g$ is the external body force (e.g., gravitational force).

Third, we need another equation to enforce the phase separation of the two fluid components and evolve their interface shapes. The Cahn-Hilliard equation can be written in terms of the relevant order parameter $C_i$:
\begin{equation}
\partial_t C_i+\mathbf{\boldsymbol{\nabla}}\cdot\left(C_i \mathbf{v}\right) = \mathbf{\boldsymbol{\nabla}}\cdot\left(M\boldsymbol{\nabla}\mu_i\right).
\label{eqn:ch}
\end{equation}
We will assume the mobility $M$ is a constant for simplicity, and the chemical potential $\mu_i = \frac{\delta \Psi}{\delta C_i}$ can be derived from the free energy of the system. 

As with the free energy minimization scheme, there are two stages of computations when employing the diffuse interface method. The first stage is to create the smooth and diffuse solid interface. This can be done by employing an energy minimization scheme as discussed in Sec.~\ref{sec.2} A. Alternatively, we can also use lattice Boltzmann approach to relax the $C_1-C_3$ (solid-gas) interface to its tanh profile.
To do this, we can write the free energy from Eq.~\eqref{frozen:total_energy} using the constraint $C_1 + C_3 = 1$,

\begin{dmath}
\Psi=\int\left[\frac{(\kappa_1+\kappa_3)}{2}C_1^2(1-C_1)^2 + \Psi_{\textrm{cf}}(\mathbf{r}) \\ 
+\frac{\alpha^2(\kappa_1+\kappa_3)}{2}(\boldsymbol{\nabla} C_1)^2+\Omega C_1^2 H(C_1)+\Omega (1-C_1)^2 H(1-C_1)\right] \textrm{d}V,
\label{Free Energy1}
\end{dmath}
where the constraining term, $\Psi_{\rm cs}$, is ignored because the lattice Boltzmann method conserves total volumes. The additional final terms using the Heaviside step function, $H$, and parametrised by
$\Omega$\cite{Lee2010} are used to prevent the order parameter deviating from the range $C_1\in(0,1)$, which can improve simulation stability during the solid initialisation when the value of $\beta$ in $\Psi_{\rm cf}(\mathbf{r})$ is large. For this initialization stage, for simplicity, we set the density, viscosity and mobility to be uniform everywhere with $\rho = 1$, $\eta = \frac{\rho}{3} (\tau-0.5) = \frac{1}{6}$ obtained by setting $\tau = 1$, and $M = \frac{(\tau_{g}-0.5)}{3} = \frac{1}{6}$ obtained by setting $\tau_{g} = 1$. As is common in lattice Boltzmann method\cite{LBM}, $\tau$ and $\tau_{\rm g}$ are relaxation parameters that are related to the fluid viscosity and mobility parameter.

Given the free energy model in Eq.~\eqref{Free Energy1}, the surface tension force and chemical potential are given by
\begin{equation}
\mathbf{F_s} = - C_1\boldsymbol{\nabla}\mu,
\end{equation}
\begin{dmath}
\mu \hiderel = 
            \frac{\delta \Psi}{\delta C_1} \hiderel = (\kappa_1+\kappa_3)C_1 (1 - 3 C_1 + 2 C_1^2)\\+\frac{\delta \Psi_{\textrm{cf}}(\mathbf{r})}{\delta C_1}-\alpha^2(\kappa_1+\kappa_3)\nabla^2 C_1+2\Omega C_1H(C_1)-2\Omega (1-C_1)H(1-C_1),
\end{dmath}
\begin{equation}
        \frac{\delta\Psi_{\textrm{cf}}(\mathbf{r})}{\delta C_1} =
        \begin{cases}
            2\beta(C_1-0.5) \quad \mathrm{if} \quad C_1 > 0.5 \quad\mathrm{and} \quad \mathbf{r} \in S, \\
             2\beta(C_1-0.5) \quad \mathrm{if} \quad C_1 < 0.5 \quad\mathrm{and} \quad \mathbf{r} \notin S, \\
             0 \quad \mathrm{otherwise}.
        \end{cases}
\label{frozen:poly-quadratic_lbm}
\end{equation}
 If the solid has sharp features or corners, to avoid numerical instabilities, we can carry out the diffuse solid initialization in two steps. First, we perform a short lattice Boltzmann run, typically with larger $\beta$ and $\Omega$ in Eq.~\eqref{Free Energy1}, and then redefine the solid locations as the positions where $C_1>0.5$. Typically, this step is only necessary where the solid geometry has corners with a radius of curvature that is comparable to $4.4\alpha$. Following this, a further initialization run is performed, where the diffuse solid is evolved using the updated solid locations. After this point, the order parameter $C_1$ is not evolved anymore.

The second stage of the simulations is to evolve the fluid order parameters ($C_2$ and $C_3$) for a fixed diffused solid ($C_1$ order parameter distribution) using the continuity, Navier-Stokes and Cahn-Hilliard equations. Starting from Eq.~\eqref{frozen:total_energy}, 
we can leave out the (now constant) terms in $C_1$ and apply the constraint that $C_3=1-C_1-C_2$ to get,
\begin{dmath}
\Psi=\int\left[\frac{\kappa_2}{2}C_2^2(1-C_2)^2+\frac{\kappa_3}{2}(1-C_1-C_2)^2(C_1+C_2)^2\\
+\alpha^2\frac{\kappa_2}{2}(\boldsymbol{\nabla} C_2)^2+\alpha^2\frac{\kappa_3}{2}(\boldsymbol{\nabla} C_1+\boldsymbol{\nabla} C_2)^2\right]\textrm{d}V.
\label{eqn:fefrozen}
\end{dmath}
The benefit of this approach is that the surface tension between fluids $C_2$ and $C_3$ and the contact angle of the $C_2$ component with the diffuse solid are given from the following relations (see Supplementary Material, Section B)
\begin{equation}
\label{eqn:st_ca}
\gamma_{23}=\frac{\alpha(\kappa_2+\kappa_3)}{6},\quad\cos\theta=\frac{\kappa_3-\kappa_2}{\kappa_2+\kappa_3}.
\end{equation}
From the definition of the free energy in Eq.~\eqref{eqn:fefrozen}, we are able to derive the surface tension force for the Navier-Stokes equation and chemical potential for the Cahn-Hilliard equation, given by 
\begin{equation}
\mathbf{F_s} = -C_2\boldsymbol{\nabla}\mu,
\end{equation}
with
\begin{eqnarray}
\mu &=& \frac{\delta \Psi}{\delta C_2} \\
&=& \kappa_2C_2 (1 - 3 C_2 + 2 C_2^2) \nonumber \\
&&+\kappa_3 (1 + C_2) (C_1^2 + 2 (-1 + C_2) C_2 + C_1 (-1 + 3 C_2)) \nonumber \\
&&-\alpha^2\kappa_2\nabla^2 C_2-\alpha^2\kappa_3\nabla^2 C_1. \nonumber
\end{eqnarray}

Now, we will give specific details for the lattice Boltzmann\cite{LBM} implementation in this study, which is used to discretise equations \eqref{eqn:ce}, \eqref{eqn:nse} and \eqref{eqn:ch}. The evolution of the macroscopic variables is expressed using distribution functions $f_i$ which are stored over a discrete regular grid. Distributions can travel to adjacent grid points in a discrete number of velocity directions $\mathbf{c}_i$ with weightings $\omega_i$ in each direction. In this study, we use the D2Q9 stencil with
\begin{equation}
\mathbf{c}=\left[\begin{array}{ccccccccc}
    0 & 1 & -1 & 0 & 0 & 1 & -1 & 1 & -1\\
    0 & 0 & 0 & 1 & -1 & 1 & -1 & -1 & 1
\end{array}\right],
\end{equation}
and
\begin{equation}
\omega=\left[\begin{array}{ccccccccc}
    \frac{4}{9} & \frac{1}{9} & \frac{1}{9} & \frac{1}{9} & \frac{1}{9} & \frac{1}{36} & \frac{1}{36} & \frac{1}{36} & \frac{1}{36}
\end{array}\right].
\end{equation}For simplicity, we choose the BGK collision operator, although extension to other collision models is possible. This choice leads to the following lattice Boltzmann equation
\begin{eqnarray}
f_{i}(\mathbf{x}+\mathbf{c}_i\delta_t,t+\delta_t)-f_{i}(\mathbf{x},t)&=&-\frac{1}{\tau}\left[f_{i}(\mathbf{x},t)-f^{eq}_{i}(\mathbf{x},t)\right] \nonumber \\
&&+\delta_t F_{i}(\mathbf{x},t).    
\end{eqnarray}
It is customary in lattice Boltzmann to take $\delta_t=1$.

Given that we solve the incompressible Navier-Stokes equation \eqref{eqn:nse}, $f_i^{eq}$ takes the form~\cite{Lee2010,IncompLBM2}
\begin{equation}
\label{eqn:feq}
f^{eq}_{i}=\omega_i\left(\frac{P}{c_s^2}+\rho s_i(\mathbf{u})\right),
\end{equation}
where $c_s^2=\frac{1}{3}$ is the sound speed and
\begin{equation}
s_i(\mathbf{u})=\left[\frac{\mathbf{u}\cdot\mathbf{c}_i}{c_s^2}+\frac{\left(\mathbf{u}\cdot\mathbf{c}_i\right)^2}{2c_s^4}+\frac{\mathbf{u}\cdot\mathbf{u}}{2c_s^2}\right]
\end{equation}
accounts for the advective terms in the Navier-Stokes equation. The form of the force $F_{i}$ is given by~\cite{Lee2010,IncompLBM2,IncompLBM3}
\begin{equation}
\label{eqn:hydroforce}
F_{i}=\left(1-\frac{1}{2\tau_f}\right)\omega_i\left[(\boldsymbol{c}_i-\boldsymbol{u})\cdot\boldsymbol{\nabla}\rho s_i(\mathbf{u})+\frac{\boldsymbol{c}_i\cdot \boldsymbol{F}}{c_s^2}\right].
\end{equation}.

Although in principle incompressibility is not vital for this method, it has been shown to reduce spurious velocities around the diffuse interface~\cite{IncompLBM2} and this form of $f^{eq}$ can aid with stability when a density ratio between fluid components included. This ensures that our model can be compatible with problems where inertia is important. Directional gradients $\boldsymbol{\nabla}$ are given by a second order isotropic central difference scheme. Moments of the distribution function provide the hydrodynamic pressure and velocity
\begin{equation}
    P=\sum_{i=0}^{8}f_i+\frac{\delta t}{2}\mathbf{u}\cdot\boldsymbol{\nabla}\rho,
\end{equation}
\begin{equation}
    \rho\mathbf{u}=\sum_{i=0}^{8}\mathbf{c}_if_i+\frac{\Delta t}{2}\mu\boldsymbol{\nabla} C.
\end{equation}

A consequence of choosing $f_i^{eq}$ from Eq.~\eqref{eqn:feq} is that the fluid density no longer arises from the distribution functions $f_i$. Instead, it is a function of the fluid concentration, which evolves through the Cahn-Hilliard equation \eqref{eqn:ch}. The lattice Boltzmann equation for the Cahn-Hilliard equation is given by
\begin{equation}
g_{i}(\mathbf{x}+\mathbf{c}_i\delta_t,t+\delta_t)-g_{i}(\mathbf{x},t)=-\frac{1}{\tau_{g}}\left[g_{i}(\mathbf{x},t)-g^{eq}_{i}(\mathbf{x},t)\right],
\end{equation}
with the equilibrium distribution $g^{eq}_{i}$ given by \cite{Zheng2015}
\begin{equation}
g^{eq}_{i}=\begin{cases}
    (1-\omega_0)\mu+C\left(1+\omega_0 s_i(\mathbf{u})\right),& \text{if } k=0,\\
    \omega_i\mu+\omega_iC s_i(\mathbf{u}),& \text{otherwise.}
\end{cases}
\end{equation}
The zeroth moment of the distribution function gives the fluid concentration
\begin{equation}
C=\sum_{i=0}^{8} g_{i}.
\end{equation}
The mobility appearing in the Cahn-Hilliard equation is given by
\begin{equation}
M=c_s^2(\tau_{g}-0.5)\delta_t.
\end{equation}
In our simulations, we relate the fluid density and viscosity as functions of the concentration as given by
\begin{eqnarray}
\rho\hiderel=\rho_3+(\rho_2-\rho_3)\frac{C_2}{C_2+C_3},\\
\tau=\tau_3+(\tau_2-\tau_3)H(C_2-C_3),\\
\eta\hiderel=\frac{\rho}{3}(\tau-0.5).
\end{eqnarray}
Moreover, in our lattice Boltzmann scheme, to implement no-slip boundary condition at the solid boundary, we apply an interpolated bounce back boundary condition \cite{Bouzidi2001}. The interface distance for the interpolation is approximated by assuming a the fluid concentration follows a $\tanh$ profile normal to the interface
\begin{equation}
\label{eqn:tanh}
C_n=\frac{1}{2}+\frac{1}{2}\tanh\frac{x}{2\alpha}.
\end{equation}
The distance to the interface along an arbitrary direction $\mathbf{\hat{d}}$ is then given by
\begin{equation}
D=\left|\frac{x}{\mathbf{\hat{n}}\cdot\mathbf{\hat{d}}}\right|,
\end{equation}
where $\mathbf{\hat{n}} = \boldsymbol{\nabla}C_n/|\boldsymbol{\nabla}C_n|$ is the normal direction to the interface. Pseudocode to describe the full simulation procedure is provided in the Supplementary Material, Section C.

\section{\label{sec.3}Phase Field Energy Minimization}
\subsection{\label{sec:level2}Liquid Contact Angle}
\begin{figure*}[ht!]
	\centering
	\includegraphics[width=\textwidth]{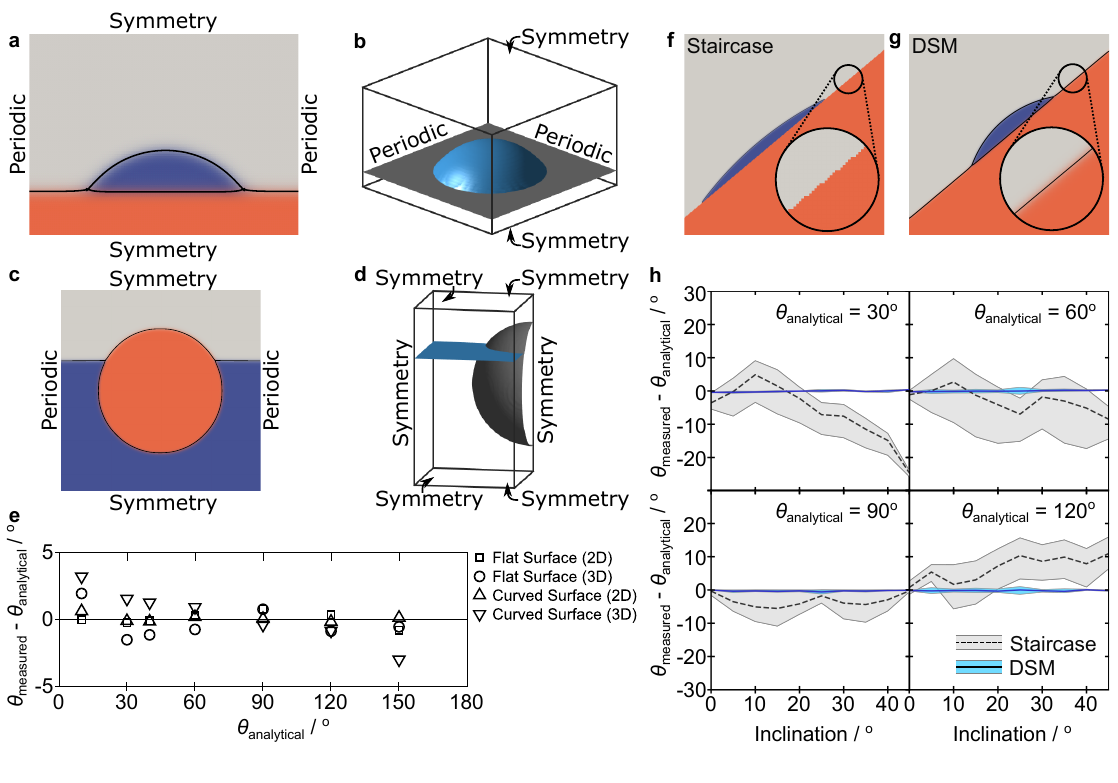}
	\caption{{Equilibrium shapes of liquids wetting (a,b) flat 
    and (c,d) curved 
    solid surfaces in 2D and 3D cases, respectively. Images also show boundary conditions used in simulations for each case. (e) Comparison of the input ($\theta_{\textrm{analytical}}$) and measured ($\theta_{\textrm{measured}}$) contact angles for wetting cases}
    on flat and curved surfaces in 2D and 3D. (f,g) Pinning investigation for a drop on an inclined plane using a binary phase field model with a staircase approximation to the surface, and using the Diffuse Solid Method (DSM) respectively. Images show a 40$^\circ$ surface inclination angle with a drop with $\theta_{\textrm{analytical}}$ = 30$^\circ$. (h) Comparison between the staircase approximation and DSM mean contact angle and range of contact angles exhibited as the volume of a drop on an inclined plane is gradually increased and reduced. Mean angles are indicated by a dashed black line (staircase) or solid black line (Diffuse Solid Method). The range of contact angles is indicated by the grey-shaded (staircase) or blue-shaded (Diffuse Solid Method) regions.} 
	\label{fig:contact_angle_test}
\end{figure*}

We first benchmark the Diffuse Solid Method by measuring the contact angle of liquid wetting on both flat and curved solid surfaces within 2-dimensional (2D) and 3-dimensional (3D) frameworks. For wetting on flat surfaces, we simulate a liquid droplet deposited on a chemically homogeneous solid surface. The simulation domain for the 2D and 3D cases are $300 \times 100$ and $300 \times 300 \times 80$, respectively. A sharp-interface solid surface of thickness 10 is initialised at the base and allowed to evolve into a diffuse interface during the minimisation process. A sessile liquid drop of radius 30 is then placed on this diffuse-interface solid, and the system is further minimized until equilibrium is achieved, yielding the final configuration shown in~\figref{contact_angle_test}(a,b). Periodic boundary conditions are applied along the horizontal directions, while symmetry boundary conditions are enforced in the vertical direction (\figref{contact_angle_test}(a,b)). 

For wetting on curved solid surfaces, we position a spherical solid particle of radius 30 at the liquid-gas interface within simulation domains of $100 \times 100$ (2D) and $60 \times 60 \times 100$ (3D). The solid particle is initialised using a staircase approximation and undergoes relaxation steps to form a smooth surface. The liquid is then introduced into the system. In the 2D case, periodic boundary conditions are applied along the horizontal directions, while symmetry boundary conditions are enforced in the vertical direction to ensure a flat liquid-gas interface, as shown in~\figref{contact_angle_test}(c). For the 3D case, computational efficiency is optimized by simulating only a quarter of the sphere, with symmetry boundary conditions imposed in all directions (\figref{contact_angle_test}(d)).


The parameters $\gamma_{23} = \sqrt{8/9}$, $\gamma_{12} = 1$, $\beta = 10$ and $n_{\text{iter}} = 10$ are used for these benchmark simulations.
The contact angles are varied between $10^\circ$ and $150^\circ$ to show a broad applicability of our method. The measurement of the contact angle is done by fitting the liquid-gas interface line, for the droplet case, or the solid interface line, for the spherical particle case, to a circle, both for the 2D and 3D cases. Note that, as shown in~\figref{contact_angle_test}(a) and (c), there is small, localized distortion at the three phase contact line. However, this deformation has negligible effect as the droplet or the particle size is always much larger, hence is ignored in the contact angle measurement.

The comparison between the input ($\theta_{\textrm{analytical}}$ from Eq.~(\ref{eqn:st_ca})) and measured ($\theta_{\textrm{measured}}$) contact angles are shown in~\figref{contact_angle_test}(e). From this figure, it can be seen that the simulation results are in good agreement with the analytical prediction for a wide range of contact angles. We note that for low contact angles ($\theta \leq 10^\circ$) in the spherical case, the three-phase contact line meets near the north pole of the sphere, and due to the diffuse interface, the accuracy of the contact angle measurement reduces, yielding a relatively large error. Otherwise, we find that the errors of the presented data are typically less than $2^\circ$. Similar accuracy for the contact angle is also obtained for the lattice Boltzmann method, as illustrated in the Supplementary Material, section D.

At this stage it is instructive to compare the accuracy of the contact angles when using the Diffuse Solid Method, with the widely-used staircase approximation. To achieve this, we perform a contact angle hysteresis test, whereby the volume of a sessile droplet is incrementally increased and then reduced, with the contact angle measured at each step. This procedure is performed on a planar surface at varying inclination angles relative to the underlying mesh, from $0^{\circ}$ to $45^{\circ}$ (all other angles being equivalent to this range). Furthermore, this test is carried out at four representative analytical contact angles, $\theta_{\textrm{analytical}}=30^{\circ}$, $60^{\circ}$, $90^{\circ}$, and $120^{\circ}$. 

In~\figref{contact_angle_test}(f) and (g), an example is shown of a 2D drop with $\theta_{\textrm{analytical}}=30^{\circ}$, on an surface of inclination angle $40^{\circ}$, for the staircase approximation and diffuse solid method respectively. Magnifications of the surfaces show the roughness introduced by the staircase approximation, compared to the smooth surface when using Diffuse Solid Method. In ~\figref{contact_angle_test}(h), the results of the contact angle hysteresis test are shown. For the staircase approximation, the mean measured contact angle $\theta_{\textrm{measured}}$, indicated by the black dashed line, deviates substantially from $\theta_{\textrm{analytical}}$ by up to $25^{\circ}$. The grey area indicates the range of $\theta_{\textrm{measured}}$ observed, which has a maximum value of $20^{\circ}$. Both of these effects are caused by the underlying surface roughness, by increasing the contact area of the drop compared to a flat plane, and introducing artificial pinning. In comparison, for the diffuse solid method, the mean $\theta_{\textrm{measured}}$, indicated by the solid blue line, deviates by a maximum of $0.4^{\circ}$, with a maximum range (light blue area) of $2^{\circ}$. Thus, the diffuse interface method is effective at creating smooth, off-lattice surface geometries.

\subsection{Capillary Rise}

\begin{figure}
	\centering
	\includegraphics[width= 7.4 cm]{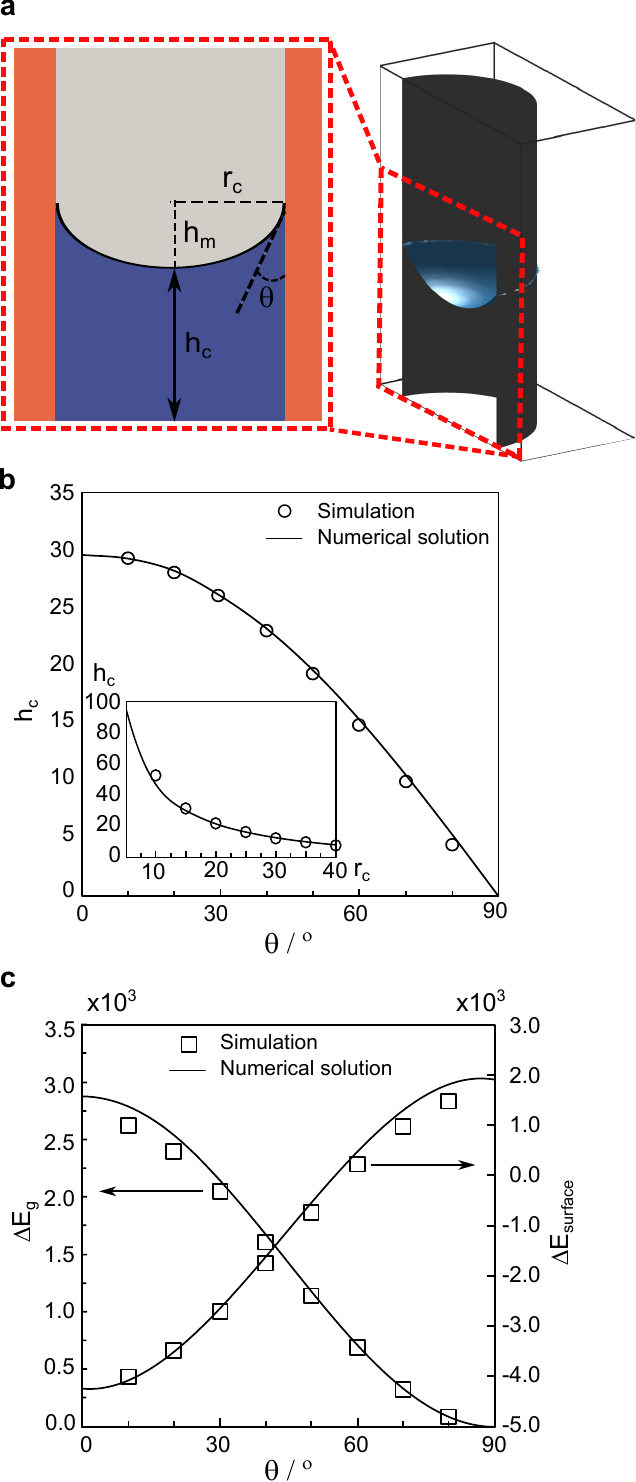}
	\caption{(a) A snapshot of the liquid-gas interface at equilibrium in a capillary rise simulation within a cylindrical tube. The tube radius $r_c$, the contact angle $\theta$, the rise height $h_c$ and the meniscus height $h_m$ are shown in the enlarged view. (b) The measured rise height $h_c$ (data points) plotted against contact angle $\theta$ with $r_c = 26$ 
    (main panel) and inner tube radii $r_c$ with $\theta = 60^\circ$ (inset), and compared to the numerical solution given by Eq.~(\ref{frozen:young-laplace}) (solid line).
    (c) The change in the gravitational potential energy $\Delta E_{\textrm{g}}$ and the surface energy $\Delta E_{\textrm{surface}}$ obtained from simulation and numerical solution plotted against contact angle $\theta$, where $r_c = 26$.
    }
	\label{fig:energy-balance-capillary-rise}
\end{figure}

We next consider the fundamental wetting phenomenon of capillary rise of a liquid in a smooth cylindrical tube~\cite{Young1805,DeGennes1985}. The equilibrium rise height of liquid within the tube can be understood from an energetic perspective. On changing the height of liquid in the tube, the total energy change of the system, $\Delta E$, can be described as a sum of interfacial ($\Delta E_{\textrm{surface}}$) and gravitational ($\Delta E_{\textrm{g}}$) contributions,
\begin{equation}
    \Delta E = \Delta E_{\textrm{surface}} + \Delta E_{\textrm{g}}. \label{frozen:total_energy_capillary_rise} 
\end{equation}
$\Delta E_{\textrm{surface}}$ can be broken into a sum of individual interfacial energy contributions, 
\begin{equation}
\Delta E_{\textrm{surface}} = \Delta A_{\textrm{lg}} \gamma_{\textrm{lg}} + \Delta A_{\textrm{ls}} \gamma_{\textrm{ls}} + \Delta A_{\textrm{gs}} \gamma_{\textrm{gs}},
\end{equation}
where $\Delta A_{\textrm{lg}}$, $\Delta A_{\textrm{ls}}$ and $\Delta A_{\textrm{gs}}$ are the area changes of the liquid-gas, liquid-solid and gas-solid interfaces respectively, and $\gamma_{\textrm{lg}}$, $\gamma_{\textrm{ls}}$, and $\gamma_{\textrm{gs}}$, are the surface tensions of the same respective interfaces. If we assume a dry capillary tube as an initial state, then we have $\Delta A_{\textrm{gs}} = - \Delta A_{\textrm{ls}} = -2\pi r_c (h_c + h_m)$, where, as shown in ~\figref{energy-balance-capillary-rise}(a), $r_c$ is the tube radius, $h_c$ is the capillary rise height (here defined from the bottom of the tube to the bottom of the meniscus) and the meniscus height $h_m$. Famously, Jurin's Law describes the equilibrium rise height in such a system, but it has only recently been shown that the precise rise height that Jurin's law describes is the mean rise height $\bar{h}$~\cite{Liu2018}:

\begin{equation}
\bar{h}= \frac{2 \gamma_{\textrm{lg}}\cos{\theta}}{\rho g r_c}.
\label{frozen:capillary_rise_height}
\end{equation}
The rise heights $h_c$ and $h_m$ cannot in general be analytically determined, and likewise neither can the energetic contributions in Eq.~(\ref{frozen:total_energy_capillary_rise}). However, as shown previously, these quantities can be determined numerically by solving the Young-Laplace equation for this axisymmetric system~\cite{Liu2018},
\begin{equation}
\frac{z''}{(1+z^{'2})^{\frac{3}{2}}}+\frac{z'}{x(1+z^{'2})^{\frac{1}{2}}}=\frac{z}{\lambda_c^2}, 
\label{frozen:young-laplace}
\end{equation}
where $\lambda_c$ is the capillary length ($\lambda_c=\sqrt{\gamma_{lg} / \rho g}$). Here, as in ref.~\cite{Liu2018}, the shooting method is employed in MATLAB to solve the meniscus profile, rise heights, and energetic contributions. 

To evaluate our simulations against this established numerical method, we begin by initialising a cylindrical tube with an inner radius of $r_c$ using a staircase approximation of a hollow cylinder with a sharp interface. We set $\beta = 10 $ in the confining potential and $n_{\textrm{iter}} = 10$ steps to smooth out the solid interface. The gravitational force is applied as a body force per unit volume, $F_b = \rho g$, in the vertical direction after the liquid phase is initialised in the system. This gravitational force enters the total free energy functional as an additional external energy term
\begin{equation}
\Psi_{\textrm{g}} = \int_V \left ( f_{\textrm{g}} C_2 z \right ) \textrm{d}V,
\label{frozen:gravitational energy}
\end{equation}
where $f_{\textrm{g}}$ is the gravitational force density. We use water as the rising liquid with properties: $\rho = 1000 $ kg $\textrm{m}^{-3}$ and $\gamma_{\textrm{lg}} = 0.0728$ N $\textrm{m}^{-1}$. The gravitational acceleration is $g = 9.81$ m $\textrm{s}^{-2}$. All of these physical parameters are then converted into simulation units via the gravitational force density: $f_{\textrm{g}} = \rho g (\gamma_{\textrm{23}} P^{'2}) / (\gamma_{\textrm{lg}} P^2)$, where $P'$ and $P$ are the length scale of the capillary tube in physical and simulation units, respectively, and $\gamma_{\textrm{lg}}$ and $\gamma_{\textrm{23}}$ are the liquid-gas interfacial tensions in physical and simulation units, respectively. We use these values in simulations: $\gamma_{\textrm{23}} = \sqrt{8/9}$, $P = 80$ and $P' = 10^{-2}$ m. To reduce the computational cost, we only simulate half of the cylinder and apply symmetry boundary conditions along all directions. The typical simulated capillary rise in a smooth cylinder tube is shown in~\figref{energy-balance-capillary-rise}(a). 

\figreff{energy-balance-capillary-rise}(b) presents the measured rise height of liquid in a smooth cylinder tube for varying wetting contact angles and tube radii (see inset). The simulations accurately capture the dependence of the rise height on $\theta$ and $r_c$, demonstrating a strong agreement with the numerical solution of Eq.~(\ref{frozen:young-laplace}). 
Additionally, we analyzed the variations in interfacial energy, $\Delta E_{\textrm{surface}}$, and gravitational potential energy, $\Delta E_{\textrm{g}}$, obtained from simulations for different contact angles, comparing them with the theoretical prediction of Eq.~(\ref{frozen:total_energy_capillary_rise}), as shown in~\figref{energy-balance-capillary-rise}(c). The change in interfacial energy is computed from the bulk and squared gradient terms within the free energy model in Eq.~(\ref{frozen:total_energy}). The change in gravitational potential energy is derived from the gravitational force term in Eq.~(\ref{frozen:gravitational energy}). The energy calculations show excellent agreement with theoretical predictions.

\subsection{Critical Pressure on Micropillared Surfaces}

We study the wetting transition on superhydrophobic surfaces, also known as the Cassie-Baxter to Wenzel transition.
In this, a liquid droplet transitions from a state suspended on top of the surface texture to a collapsed state where the droplet fully wets the structure. This is associated with a reduction in the apparent contact angle of the droplet and a loss of superhydrophobicity. The wetting transition occurs if the pressure difference between the liquid and gas is larger than a critical pressure given by the surface textures, which depends upon the equilibrium contact angle. At higher equilibrium contact angles, it is harder for the liquid to enter the texture, resulting in a higher critical pressure.

\begin{figure}
	\centering
	\includegraphics[width=8cm]{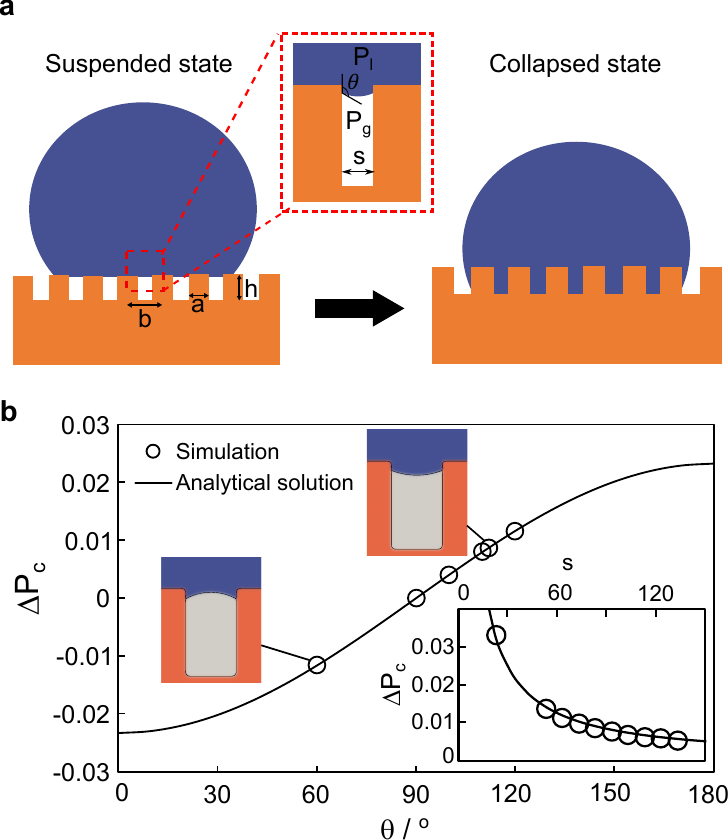}
	\caption{(a) Schematic of a droplet undergoing a transition from a suspended state to a collapsed state on a 2-dimensional periodically patterned surface with a rectangular micropillar geometry having width \textit{a}, pillar spacing \textit{s}, pillar height \textit{h} and periodicity \textit{b}. 
    $\theta$ is the equilibrium contact angle on the corresponding flat surface. $P_{\textrm{l}}$ and $P_{\textrm{g}}$ are the pressure in the liquid and gas (surrounding air), respectively. The enlarged view shows the liquid-gas interface underneath the droplet hanging between two adjacent pillars in a suspended state and is used as the system setup in simulations.
    (b) Measured critical pressure plotted against contact angle, where $s = 81$, and pillar spacing (inset), where $\theta = 112^\circ$, and compared with the analytical prediction to Eq.~(\ref{frozen:critical-pressure-cyl-2d}). Snapshots show liquid-gas interface shapes at the critical pressure for $\theta=60^\circ$ and $\theta=112^\circ$. }
	\label{fig:droplet-cassie-baxter}
\end{figure}

We consider a 2D system where the solid is textured with rectangular micropillars, as shown in~\figref{droplet-cassie-baxter}(a). We simulate a liquid layer suspended on top of two pillars separated by a gap \textit{s} instead of simulating a whole droplet on a patterned surface (see enlarged view in~\figref{droplet-cassie-baxter}(a)). We make this approximation assuming that the droplet size is much larger than the pillar size and spacing. In our simulations, these values are used: $a = 40$, $h = 170$, $b=160$, $\beta = 10$ and $n_{\text{iter}} = 10$. The system is fitted into a simulation box of $160 \times 200$. We incorporate pressure into the system by applying a pressure difference between the liquid and gas phase, $\Delta P$, in the free energy functional in Eq.~(\ref{frozen:pressure_const}). 
In order to obtain an accurate critical pressure, we apply a binary search algorithm~\cite{williams1976modification,knuth1997art} around the critical pressure value with a tolerance level of $10^{-4}$. 

Theoretically, for the 2D model system used here, the critical pressure is given by \cite{mitra2017wetting,patankar2010consolidation,giacomello2012metastable}
\begin{equation}
\Delta P_c = -\frac{2\gamma_{\textrm{lg}}\cos{\theta}}{s},
\label{frozen:critical-pressure-cyl-2d}
\end{equation}
where 
$\gamma_{\textrm{lg}}$ is the liquid-gas surface tension; $\theta$ is the equilibrium contact angle of the droplet on the corresponding flat surface, and $s = b-a$ is the wall-to-wall spacing between two adjacent pillars. 

The equilibrium shapes of the liquid-gas interface at the critical pressure are shown in the snapshots of~\figref{droplet-cassie-baxter}(b), in which concave and convex menisci occur due to the liquid-solid surface tension for $\theta = 60^\circ$ and $\theta = 112^\circ$, respectively. At the transition ($\Delta P > \Delta P_c$), the suspended liquid-gas interface abruptly collapses, enabling the liquid phase to displace the trapped gas and completely fill the gap between the pillars. The critical pressure dependency on the equilibrium contact angle, $\theta$,  and the pillar spacing, \textit{s}, is shown \figref{droplet-cassie-baxter}(b). As predicted by Eq.~(\ref{frozen:critical-pressure-cyl-2d}), the critical pressure is linearly dependent on $\cos{\theta}$. For $\theta < 90^\circ$, the critical pressure is negative, indicating that the energy barrier is low and the transition to the collapsed state is energetically favourable. On the contrary, the positive critical pressures for $\theta > 90^\circ$ suggest the system has a higher energy barrier, making the transition not energetically favourable to occur unless some external perturbations from the liquid pressure are added to the system. As the separation increases, the critical pressure reduces making it more favorable for the liquid to transition to a collapsed state. The simulated critical pressures are in good agreement with the prediction. 

\subsection{The Salvinia Paradox}

\begin{figure*}
	\centering
	\includegraphics[width=\textwidth]{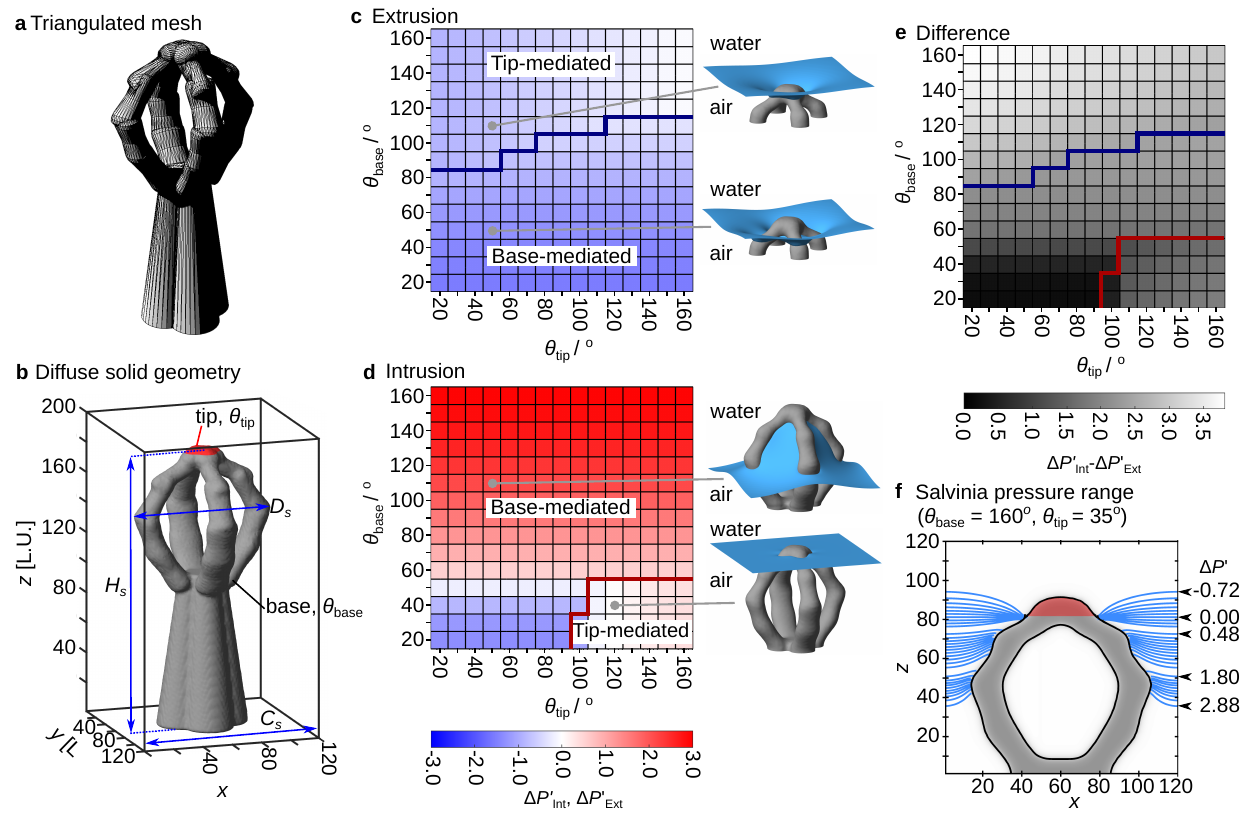}
	\caption{(a) The triangulated 3D mesh that models the 'eggbeater' shaped leaf hair of the water-fern, \textit{Salvinia molesta}. (b) The diffuse solid conforming to the inputted triangulated mesh, showing the grid dimensions and different contact angles imposed on the base, $\theta_{\rm{base}}$, and the tip, $\theta_{\rm{tip}}$.Key hair parameters are indicated: the hair height $H_{\mathrm{s}}$, outermost diameter $D_{\mathrm{s}}$, and centre-to-centre distance $C_{\mathrm{s}}$.  (c-e) Surveys of the nondimensionalised extrusion pressure $\Delta P'_{\rm{Ext}}$, intrusion pressure $\Delta P_{\rm{Int}}$, and difference between these pressures respectively. Boundaries between the two collapse modes (Tip-mediated and Base-mediated) are indicated by a thick blue line (extrusion) and thick red line (intrusion). (f) A 2D slice through the \textit{Salvinia} hair, showing how the liquid-gas interface changes across the pressure range between the intrusion and extrusion pressures. Interfaces are shown in blue at equal increments of $\Delta P'$ = 0.12. The hydrophilic tip is shown in red, the hydrophobic base in grey.}
	\label{fig:salvinia}
\end{figure*}

Having validated the Diffuse Solid Method, we now demonstrate its utility in representing real structures of complex geometry through a case study of the so-called 'Salvinia Paradox'. First investigated by Barthlott \textit{et al.} in 2010 \cite{Barthlott2010}, the Salvina Paradox concerns the wetting properties of the hair-like structures that coat the leaves of the \textit{Salvina} water fern. In the species \textit{Salvina molesta}, these mm-long hairs have a complex, 'eggbeater' geometry, which is modelled in \figref{salvinia}(a). The habitat of \textit{Salvina molesta} is on the surface of rivers and lakes, and plants are susceptible to being submerged. However, the plants also require an air layer close to the leaf (plastron) to be maintained gaseous exchange. The eggbeater hairs are an adaptation that allow the plastron to be maintained when submerged. The four arms are coated in a hydrophobic wax, and through their physical structure maintain a superhydrophobic suspended state when submerged. This resists the intrusion of water into the plastron. The paradox arises because the uppermost tip of the hair is hydrophilic, a feature which conventionally is assumed to be detrimental to superhydrophobicity. However, it has been shown experimentally that whereas the hydrophobic structured base resists the intrusion of water, the hydrophilic tip resists the exit of air as bubbles from the plastron. 
These two different wetting behaviours are suggested to work synergistically to maintain the plastron, even in flowing water \cite{Barthlott2010}.

To measure the effectiveness of these structures at resisting breakdown of the plastron, it is useful to consider the extrusion and intrusion pressures, the pressures at which the liquid-gas interface either detaches upwards from the tip of the structure, or moves spontaneously down the structure towards the base respectively. To date, these properties have largely been investigated experimentally on plant samples, with reported extrusion pressures of (-6.1 $\pm$ 0.9) kPa \cite{Gandyra2020}, and intrusion pressures of (12 $\pm$ 2) kPa \cite{Mayser2014}. However, there remains a lack of fundamental understanding about the relationship between the structure, surface wettability, and the extrusion/intrusion pressures. Towards achieving this understanding, some progress has been made via computational studies. The extrusion and intrusion characteristics on a 2D simplified geometry showed how the interaction between wettability patterning and structure enhanced the difference between extrusion and intrusion pressures \cite{Amabili2015}. Meanwhile more recently, the apparent contact angle of the suspended state was investigated, alongside some aspects of the intrusion characteristics using the proprietary software Ansys Fluent \cite{Ansys}, however wettability patterning was not considered \cite{Zhang2023}. 

Using the Diffuse Solid Method, the extrusion and intrusion of water on the complex geometry with patterned wettability is investigated.
To begin with, a model of the geometry was designed in Blender \cite{Blender2024} (although any suitable modelling or computer-aided design (CAD) software is appropriate). The triangulated mesh of this model is shown in \figref{salvinia}(a).
Within the computational domain of the phase field model, the face and vertex data of the triangulate mesh was then used to distinguish the solid interior and exterior, and so set up the confining potential $\Psi_{\textrm{cf}}$ described in Eq.~(\ref{frozen:poly-quadratic}). Following the solid relaxation step, the resultant solid geometry is shown in \figref{salvinia}(b), in which the hydrophobic base of contact angle $\theta_{\rm{base}}$ is shown in grey, and the hydrophilic tip of contact angle $\theta_{\rm{tip}}$ is highlighted in red.

When choosing the dimensions of the model hair and domain size, it is recognised that real plant specimens show substantial variability in the hair geometry and arrangement \cite{Ditsche2015}. Key geometric parameters commonly examined are the hair height $H_{\mathrm{s}}$, outermost diameter $D_{\mathrm{s}}$, and centre-to-centre distance $C_{\mathrm{s}}$, indicated in \figref{salvinia}(b). Here, for this demonstration of the Diffuse Solid Method, we opted for a typical hair geometry characterised by 
$D_{\mathrm{s}}/C_{\mathrm{s}}$ = 0.77, and  $H_{\mathrm{s}}/C_{\mathrm{s}}$=0.99.  Periodic boundary conditions were imposed in the \textit{x} and \textit{y} directions, with symmetry boundary conditions in the \textit{z} directions.  

To examine the extrusion pressure, $\Delta P_{\rm{Ext}}$, a liquid layer was initialised in contact with the tip of the hair, and the energy of the system minimised. The pressure in the liquid phase was reduced relative to the gas phase and the system energy re-minimised. This step was repeated with variable steps in the pressure changes to locate the pressure at which the interface detached from the top of the structure: $\Delta P_{\rm{Ext}}$. To explore the impact of the base and tip wettability on $\Delta P_{\rm{Ext}}$, this process was repeated across a range of $\theta_{\rm{base}}$ and $\theta_{\rm{tip}}$ from $20^{\circ}$ to $160^{\circ}$, capturing the range of biologically relevant contact angles.

The variation of $\Delta P_{\rm{Ext}}$ with $\theta_{\rm{base}}$ and $\theta_{\rm{tip}}$ is shown in \figref{salvinia}(c), with the color scale located at the bottom of the figure. Note that $\Delta P'_{\rm{Ext}}$ is reported, which is a nondemensionalised form of $\Delta P_{\rm{Ext}}$: $\Delta P'_{\rm{Ext}}=\Delta P_{\rm{Ext}} / (\gamma_{\rm{lg}}/L_{\rm{x}})$, where $\gamma_{\rm{lg}}$ is the liquid-gas interfacial tension ($\gamma_{\rm{lg}}=1.0$) and $L_{\rm{x}}$ is the number of cells in the computational domain in the $x$-direction ($L_x=120$). Two different modes of interface detachment are observed, labelled Base-mediated and Tip-mediated, which are separated by the thick, blue line. In Base-mediated detachment, at the point of detachment, the interface is pinned to the solid base, and so $\theta_{\rm{tip}}$ does not effect $\Delta P_{\rm{Ext}}$. Instead, the smaller the value of $\theta_{\rm{base}}$, the more negative $\Delta P_{\rm{Ext}}$ becomes. Conversely, in Tip-mediated detachment, the interface is pinned to the tip at the point of detachment and so $\Delta P_{\rm{Ext}}$ becomes independent of $\theta_{\rm{base}}$. The smaller the value of $\theta_{\rm{tip}}$, the more negative $\Delta P_{\rm{Ext}}$ becomes. However, the small perimeter of the three-phase contact line in Tip-mediated detachment limits the maximum magnitude of $\Delta P_{\rm{Ext}}$. This is compared to Base-mediated detachment where the relatively large three-phase contact line perimeter enables larger magnitude extrusion pressures to be achieved. Thus, if preventing extrusion was the sole purpose of the \textit{Salvinia} hairs, then this would most effectively be achieved by minimising $\theta_{\rm{base}}$.

To measure the intrusion pressures, $\Delta P_{\rm{Int}}$, the pressure in the liquid relative to the gas was iteratively increased up until the point where the liquid spontaneously filled the system. We note as an aside that depending on $\theta_{\rm{base}}$, the interior of the my either trap an air bubble, or fill with liquid. The variation of $\Delta P'_{\rm{Int}}$ with $\theta_{\rm{base}}$ and $\theta_{\rm{tip}}$ is shown in \figref{salvinia}(d), where $\Delta P'_{\rm{Int}}$ has been nondimensionalised in an identical manner to $\Delta P'_{\rm{Ext}}$. Here, we also see two separate detachment mechanisms, Base-mediated and Tip-mediated. In Base-mediated detachment, the three-phase contact line pins close to the widest part of the hair geometry at the point of detachment. The eggbeater structure traps air within the space between the four arms, making it energetically expensive for the liquid-gas interface to move down the outside of the hair. Thus, the intrusion pressure is positive, even for hydrophilic values of $\theta_{\rm{base}}$ as low as $50^{\circ}$. In scenarios where the base is hydrophilic and the tip highly hydrophobic (the reverse of that seen in the \textit{Salvinia} plant), the tip becomes responsible for resisting intrusion in Tip-mediated detachment. However, due to the small contact line perimeter of the tip, only modest positive intrusion pressures can be achieved. If preventing intrusion was the sole purpose of the \textit{Salvinia} hairs, then this would most effectively be achieved by maximising $\theta_{\rm{base}}$.

To explore the proposed function of the wettability-patterened \textit{Salvinia} hair (to maintain the largest range between intrusion and extrusion pressures \cite{Barthlott2010}), we show the difference, $\Delta P_{\rm{Int}}$-$\Delta P_{\rm{Ext}}$, in \figref{salvinia}(e). This shows that, in order to maximise $\Delta P_{\rm{Int}}$-$\Delta P_{\rm{Ext}}$, $\theta_{\rm{base}}$ should be maximised, and $\theta_{\rm{tip}}$ minimised. Although not yet explicitly studied for salvinia, the nano-structured wax coating of the hydrophobic base is anticipated to show similar contact angles to other plants, approximately $160^{\circ}$ \cite{Barthlott2010,Barthlott1997}. Likewise, although not yet robustly confirmed, the hydrophilic tip is suggested to have a contact angle of approximately $35^{\circ}$ \cite{Gandyra2020}. At these contact angles, our simulations show that $\Delta P'_{\rm{Ext}}$ = -0.81, $\Delta P'_{\rm{Int}}$ = 2.94, with a difference of 3.75. Choosing a typical value of the centre-to-centre hair spacing of 790 $\mu$m \cite{Ditsche2015}, these convert to $\Delta P_{\rm{Ext}}$ = -8.85 kPa, $\Delta P_{\rm{Int}}$ = 32.1 kPa, with a difference of 41.0 kPa. This compares sensibly to the experimental values of $\Delta P_{\rm{Ext}}$ = (-6.1 $\pm$ 0.9) kPa \cite{Gandyra2020} and $\Delta P_{\rm{Int}}$ = (12 $\pm$ 2) kPa \cite{Mayser2014}. Note that the magnitudes of these pressures are reduced compared to the simulated values. This is because in real plant samples, substantial geometric variability between hairs means that measured detachment pressures will be dominated by the local regions which are least resistant to intrusion or extrusion. In the Supplementary Material (Section E), a sensitivity analysis is performed to show that the intrusion and extrusion critical pressures are highly sensitive to $D_{\mathrm{s}}/C_{\mathrm{s}}$, and only moderately sensitive to $\theta_{\rm{base}}$ and $\theta_{\rm{tip}}$. Finally, an interesting further analysis is that if the hair exhibited a uniform contact angle, the maximum extrusion-intrusion pressure difference occurs at $\theta_{\rm{base}}=\theta_{\rm{tip}}$ = $160^{\circ}$, with a value of 32.6 kPa. Thus, for the geometry tested here, wettability patterning increased the plastron stability pressure range by 26$\%$.

To further illuminate the \textit{Salvinia} hair function, in  \figref{salvinia}(f), a 2D slice through the 3D hair tip is shown. Between the intrusion and extrusion pressures, the liquid-gas interface is shown at uniform intervals of $\Delta P'$ = 0.12. At zero pressure, the interface is pinned on the hydrophobic/hydrophilic boundary. If the pressure is made increasingly negative (extrusion conditions), it can be seen that the three-phase contact line remains pinned to this location up until $P_{\rm{Int}}$ is reached. If instead the pressure is increased gradually from 0 (intrusion conditions), at $\Delta P'$ = 0.48, the three-phase contact line depins and slides partially down the hair. As the pressure is increased further, the contact line continuously slides down the hair until $\Delta P'$ = 1.80, at which point another jump in contact line location is observed. From here, the interface slides towards the widest part of the hair, up until $P_{\rm{Ext}}$ is reached.

\section{\label{sec.4}Lattice Boltzmann Simulations} 

\subsection{Capillary Filling Dynamics}

In order to benchmark the accuracy of the diffuse solid for dynamic problems, we investigate the filling of a capillary channel\cite{capfill}. 
The setup involves a finite reservoir of liquid entering a narrow channel, which the liquid preferentially wets over the gas. The curved liquid-gas interface in the channel creates a pressure gradient between the liquid and the surrounding gas, which subsequently drives the liquid to fill the channel. In the case where the liquid is much more viscous than the gas, the distance traveled by the liquid into the channel follows a square-root dependence on time
\cite{washburn}
\begin{equation}
L-L_0=\left(\frac{\gamma_{\rm{lg}}H \cos\theta_{\rm d}}{3\eta_{\textrm{l}}}\right)^{\frac{1}{2}}(t-t_0)^{\frac{1}{2}}.
\label{eqn:cfilling}
\end{equation}
Here, $\eta_{\textrm{l}}$ is the dynamic viscosity of the liquid, $\theta_{\rm d}$ is the dynamic contact angle, $\gamma_{\rm{lg}}$ is the liquid-gas interfacial tension, $H$ is the channel width, and $t$ is time. $L_0$ and $t_0$ are constants of integration that take into account the initial dynamics of the liquid as it enters the channel.

\begin{figure}[h!]
	\centering
	\includegraphics[width=8cm]{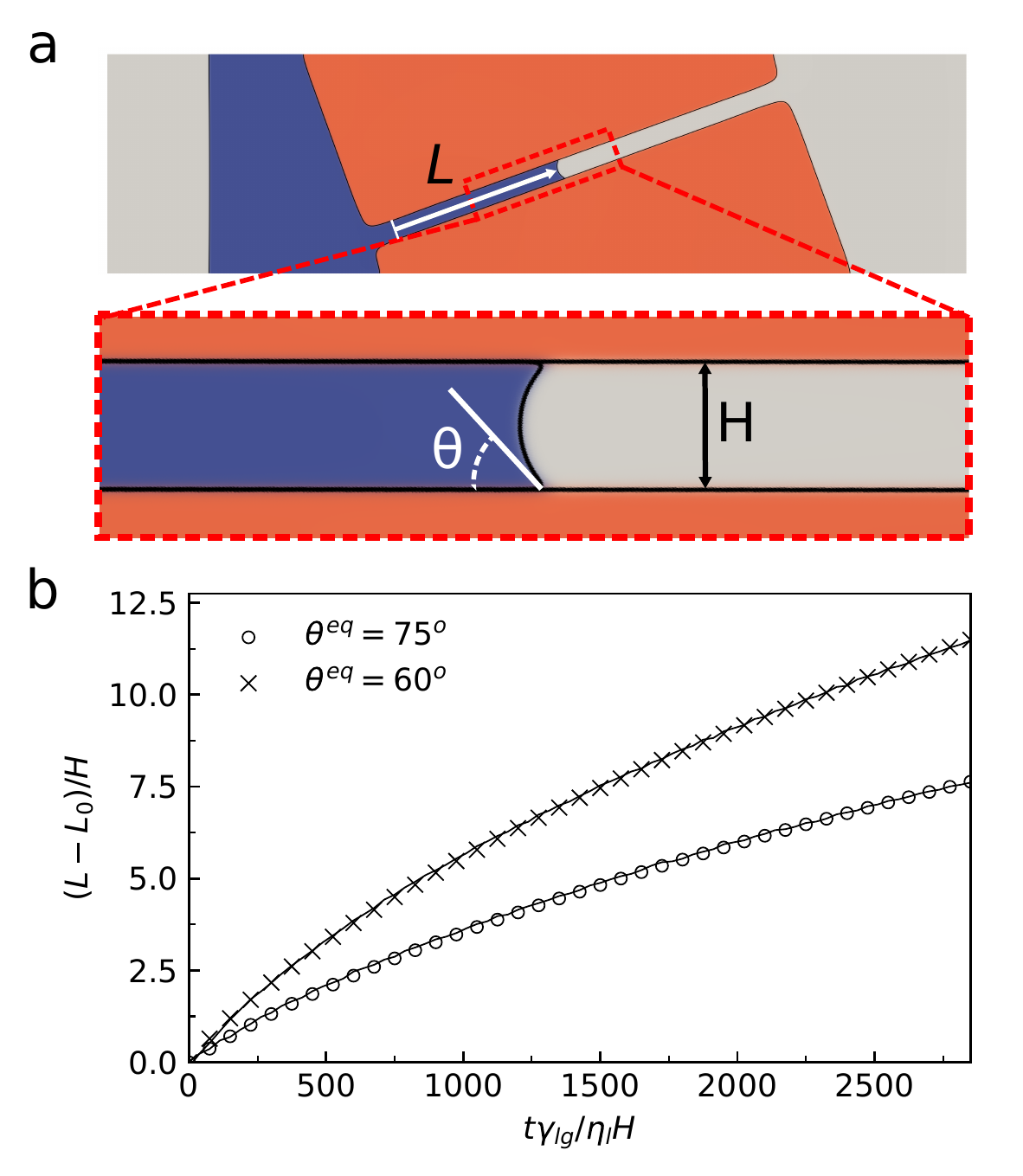}
	\caption{(a) Visualisation of simulation domain, with a reservoir of liquid on the left, which will enter the blue solid structure. The channel is inclined at $20^\circ$. The boundaries on the left and right are periodic with symmetry conditions on the top and bottom \cite{LBM}. (b) Simulation data points and analytical curves (shown by solid lines), for non-dimensional distance and time.}
	\label{fig:capillary_filling}
\end{figure}

\figref{capillary_filling} shows how the distance traveled by the liquid into the channel depends on time. The simulation parameters are: a surface tension 
$\gamma_{lg}=2.5 \times 10^{-3}$, an interface width $\alpha=1$ and viscosities of $1/6$ and $2/300$ for the liquid and gas phases respectively. The densities of each component are kept identically at $1$. The value of mobility $M$ in the Cahn-Hilliard equation is $1/6$. The solid is initialised by equilibrating for 
$5 \times 10^3$ iterations using 
$\beta = 1 \times 10^{-3}$ and $\Omega = 2 \times 10^{-4}$.

The full simulation size is $1750 \times 450$ and the height and width of the channel are $900$ and $40$ respectively. We incline the channel by $20^\circ$ in order to validate the dynamics in the case that the diffuse solid does not align to the grid. 

These simulations are run for equilibrium contact angles of $60^\circ$ and $75^\circ$. However, as expected from previous literature \cite{capfill,capfill2}, the dynamic contact angles are larger than these values. The dynamic contact angle is measured by fitting a circle to points on the liquid-gas interface and finding the tangential angle at the point where this intercepts the solid plane. As the liquid front moves along the channel and the velocity drops, the measured contact angles approach their equilibrium values, see Supplementary Material (section F), providing another validation that we achieve the correct equilibrium wetting behaviour from Eq.~(\ref{eqn:fefrozen}). 

In \figref{capillary_filling}(b), the time and distance are non-dimensionalised by choosing
\begin{equation}
L^*=L/H,\quad t^*=t\gamma_{\textrm{lg}}/\eta_{\textrm{l}} H.
\end{equation}
We fit the data to Eq.~(\ref{eqn:cfilling}) using a least squares algorithm, using the instantaneous dynamic contact angle at each data point when fitting the curve.  As shown by \figref{capillary_filling}(b), the distance travelled by the liquid interface follows a square root dependence on time. The fit agrees well with the expected curve given by Eq.~(\ref{eqn:cfilling}). These results provide confidence that the capillary dynamics are captured accurately. The curved interface provides the correct pressure gradient to drive the capillary flow with the expected velocity for a large viscosity contrast.

\subsection{Droplet Dynamics on Sinusoidal Surfaces}

To highlight a complex dynamic wetting application of the diffuse solid method, we now demonstrate simulations of droplets moving on sinusoidally corrugated surfaces.
The solid initialization approach enables the generation of arbitrary solid shapes. This capability is vital to simulate multicomponent flow on real geometries, which very rarely conform to a regular simulation grid. Sinusoidal surfaces are a key example for this, and have found growing interest as superhydrophobic surfaces due to their unique dynamic behaviour. In contrast to traditional pillared surfaces, the smooth texture limits pinning\cite{sinusoidal}, so droplets can 
move more easily. Additionally, these wavy surfaces more closely match biological superhydrophobic surfaces, which have shown remarkable capabilities to repel liquid droplets\cite{Cheng20061359,sinusoidal2}. 

\begin{figure}[h!]
	\centering
	\includegraphics[width=8cm]{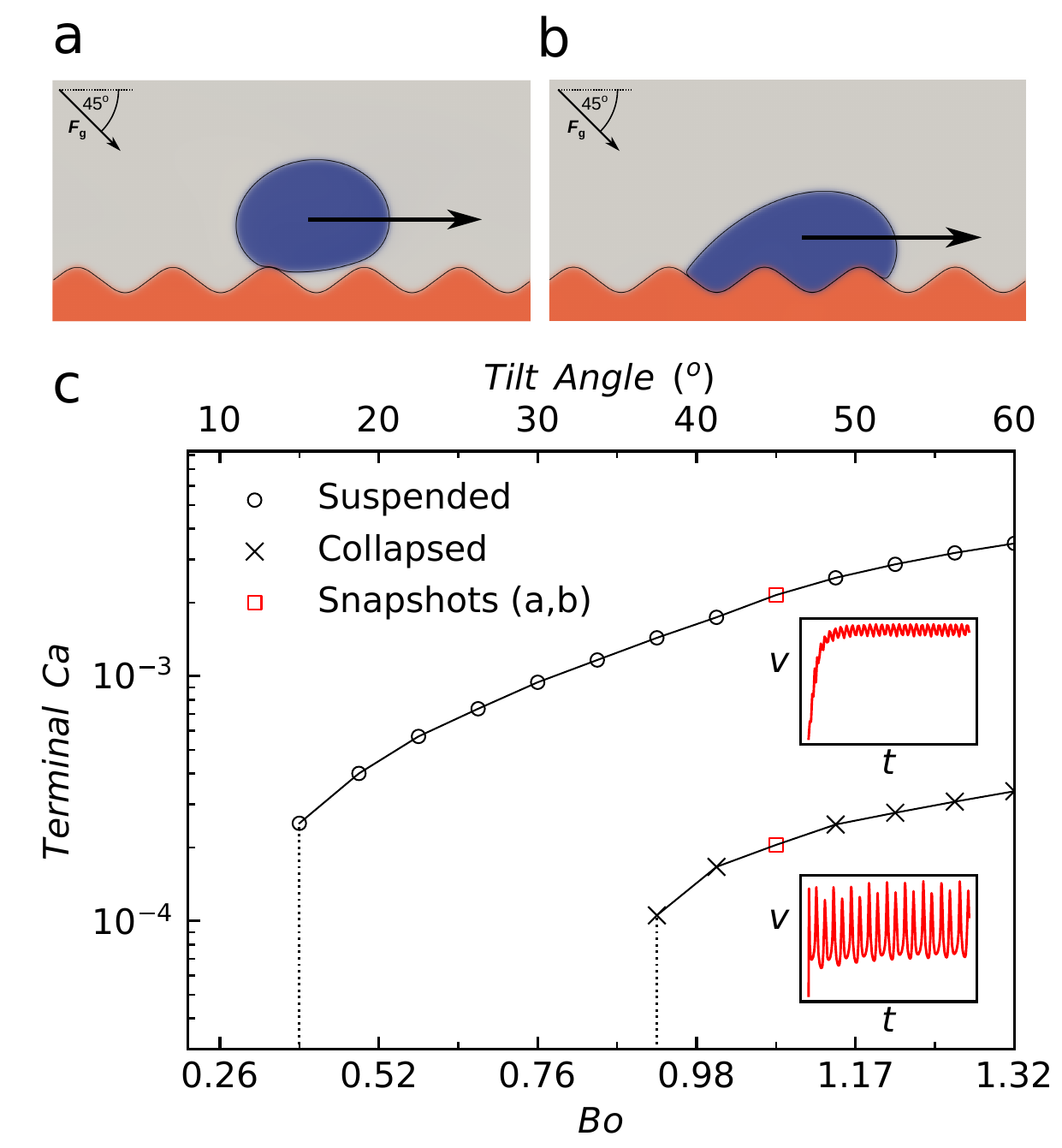}
	\caption{ (a) Droplet in the suspended state at a tilt angle of $45^\circ$, raised above the corrugated texture as it moves. The equilibrium contact angle is $150^\circ$. (b) Droplet in the collapsed state at a tilt angle of $45^\circ$ that has penetrated the texture. The equilibrium contact angle is $90^\circ$. (c) Terminal capillary number as a function of Bond number/Tilt Angle. Insets show that the velocity is close to constant in the suspended state and exhibits stick-slip motion in the collapsed state. The droplet in the suspended state moves at a significantly smaller tilt angle of $15^\circ$ compared with $37.5^\circ$ for the collapsed state. In addition to this, the terminal capillary number is consistently larger by approximately an order of magnitude.}
	\label{fig:sin_surface}
\end{figure}

We investigate a droplet moving on these surfaces in the cases where the droplet is in the suspended and collapsed states, shown in \figref{sin_surface}(a-b). 
We initialise the solid component following a sinusoidal profile with an amplitude of $20$ and a period of $50$. We then allow the interface to smooth over 2500 timesteps with 
$\beta = 2 \times 10^{-3}$ and $\Omega = 0.1$, followed by a further 47500 timesteps with $\beta = 2 \times 10^{-3}$ and $\Omega = 2 \times 10^{-3}$.
To achieve the 
suspended and collapsed states, we set the equilibrium contact angles to $150^\circ$ and $90^\circ$ respectively. The other simulation parameters are identical to the values in Sec.~\ref{sec.4} A. The full simulation size is $250 \times 130$ and each simulation was run for 
$2 \times 10^6$ timesteps. Periodic boundary conditions are used at the left and right of the domain, and the halfway bounce-back scheme\cite{LBM} is used at the top. At the beginning of the simulation, a circular droplet with a radius of $40$ is initialised in the collapsed state. To apply gravity, we include a constant body force to the droplet (we neglect gravity on the surrounding component with the assumption that the density ratio is large). The magnitude of the body force is calculated to match the Bond number 
\begin{equation}
\textrm{Bo}=\frac{(\rho_{\textrm{water}}-\rho_{\textrm{air}})D^2g*}{\gamma_{\textrm{water,air}}}
\end{equation}
of a water droplet with a diameter $D=3.37$ mm. Here $g*$ is the gravitational acceleration parallel to the surface.
The terminal velocity $V$ is calculated as the average center of mass velocity parallel to the solid surface for the final 
$1 \times 10^6$ timesteps, after the variation in average velocity of the droplet over successive periods of the sinusoidal texture was below $1\%$. 

The results of these simulations is presented in \figref{sin_surface}(c), showing the terminal capillary number
\begin{equation}
\textrm{Ca}=\frac{\eta_{\textrm{l}} V}{\gamma_{\textrm{lg}}}.
\end{equation}
These results demonstrate the abilities of sinusoidal surfaces to repel liquid droplets. If the underlying substrate is hydrophobic, the droplet will enter the suspended state, and easily roll over the top of the corrugated texture with a small tilting angle of $15^\circ$. Even while the surface is neutrally wetting
($\theta=90^\circ$),
the collapsed state still begins to move at a moderate tilt angle of $37.5^\circ$, due to the lack of sharp corners on the sinusoidal surface, unlike other superhydrophobic surfaces patterned with posts.

Insets in \figref{sin_surface}(c) display the typical velocity variation with time for the suspended and collapsed states. For the suspended state, the velocity stabilises quickly, with small oscillations as the droplet passes the sinusoidal profile. In contrast, for the collapsed state, the droplet must move vertically to overcome the peaks of the sinusoidal profile, which leads to a much smaller velocity with large oscillations as the droplet moves from one sinusoidal period to the next. The alternating pattern in the peak velocity of these oscillations arises from the different curvatures of the front and rear of the droplet in \figref{sin_surface}(b). Consequently, this leads to a slightly larger velocity as the front of the droplet overcomes the texture, in contrast with the rear. From these results, we also observe that the diffuse solid approach allows these surfaces to be fully resolved without a staircase-like approximation. The latter would create artificial pinning locations. Moreover, the solid initalization method allows for the correct interface profile to be achieved when analytical smoothing is impossible, which is essential for wetting on arbitrary geometries.


\section{\label{sec.5}Conclusion}

In this work we have presented a diffuse solid method that allows us to study quasi-static and dynamic wetting and multiphase flow phenomena in complex geometries. The diffuse solid approach is simple to implement and can be embedded in different numerical solvers, as shown here for phase field energy minimization and lattice Boltzmann method. Moreover, we have obtained excellent agreement against analytical results for all the benchmark scenarios;
equilibrium contact angles, capillary rise height, critical pressure on pillared surfaces, and dynamic capillary rise.
In turn, the diffuse solid model allowed us to easily study phenomena involving highly non-trivial geometries, highlighted here using the Salvinia leaf structure with its eggbeater shape and variation in contact angle at the base and the tip as well as droplet motion on sinusoidal surfaces. 

We believe the proposed diffuse solid method will open up new opportunities to investigate wide ranging interfacial phenomena. For instance, it is useful for rationalizing wetting dynamics on bio-inspired surfaces, ranging from Lotus leaves and Rose petals to springtails and butterfly wings~\cite{Chatenet2024355,Mehrjou2023,Oopath2023,Jin20231620}, which have been shown to exhibit a rich display of variations in shapes, sizes, and chemical variations. It can also be harnessed as a powerful design tool, especially when coupled with CAD software (as has been done here for the Salvinia leaf study), and potentially 3D printing to experimentally realize the computationally optimized surface structures.

\section*{\label{sup.mat}Supplementary material}

The Supplementary Material contains additional information on (A) a parameter sensitivity study to obtain specific solid geometries, (B) the derivation of interfacial tensions and contact angle in the Diffuse Solid Model, (C) pseudocode for the lattice Boltzmann algorithm emloyed in this work, (D) equilibrium contact angle validation for the lattice Boltzmann method simulations, (E) parameter sensitivity survey for the {\it Salvinia paradox} example, and (F) dynamic contact angle data for the lattice Boltzmann simulations shown in Sec. IVB.

\begin{acknowledgments}

F.O. acknowledges a BPPLN scholarship from the Directorate General for Science Technology and Higher Education, Republic of Indonesia. H.K. acknowledges funding from the Leverhulme Trust (Research Project Grant RPG-2022-140) and UKRI Engineering and Physical Sciences Research Council (EP/V034154/2). H.K. and S.J.A. acknowledge funding from EPSRC Impact Acceleration Account and ExxonMobil.

\end{acknowledgments}

\section*{Author Declarations}
\subsection*{Conflict of Interest}
The authors have no conflict to disclose.
\subsection*{Author Contributions}
\textbf{Fandi Oktasendra:} Conceptualization (equal); Investigation (equal); Methodology (equal); Writing - Original Draft (equal). \textbf{Michael Rennick:} Investigation (equal); Methodology (equal); Writing - Original Draft (equal). \textbf{Samuel J. Avis:} Methodology (equal); Writing - Review \& Editing (equal). \textbf{Jack R. Panter:} Conceptualization (equal); Investigation (equal); Methodology (equal); Writing - Original Draft (equal). \textbf{Halim Kusumaatmaja:} Conceptualization (equal); Funding acquisition (lead); Supervision (lead); Writing - Original Draft (equal).

\section*{Data Availability Statement}
Data supporting the findings in this study are available from the corresponding authors upon reasonable request. 


\clearpage

\onecolumngrid

\section{Supplementary Material}

\subsection{Diffuse Solid Method Parameter Survey}

We have carried out similar surveys for both the phase field energy minimization and lattice Boltzmann simulations to optimize the choice of parameters for the Diffuse Solid Method. For brevity, here we present the results for the phase field energy minimization. 

\subsubsection{Solid confinement parameters}

\begin{figure}
	\centering
	\includegraphics[width=16cm]{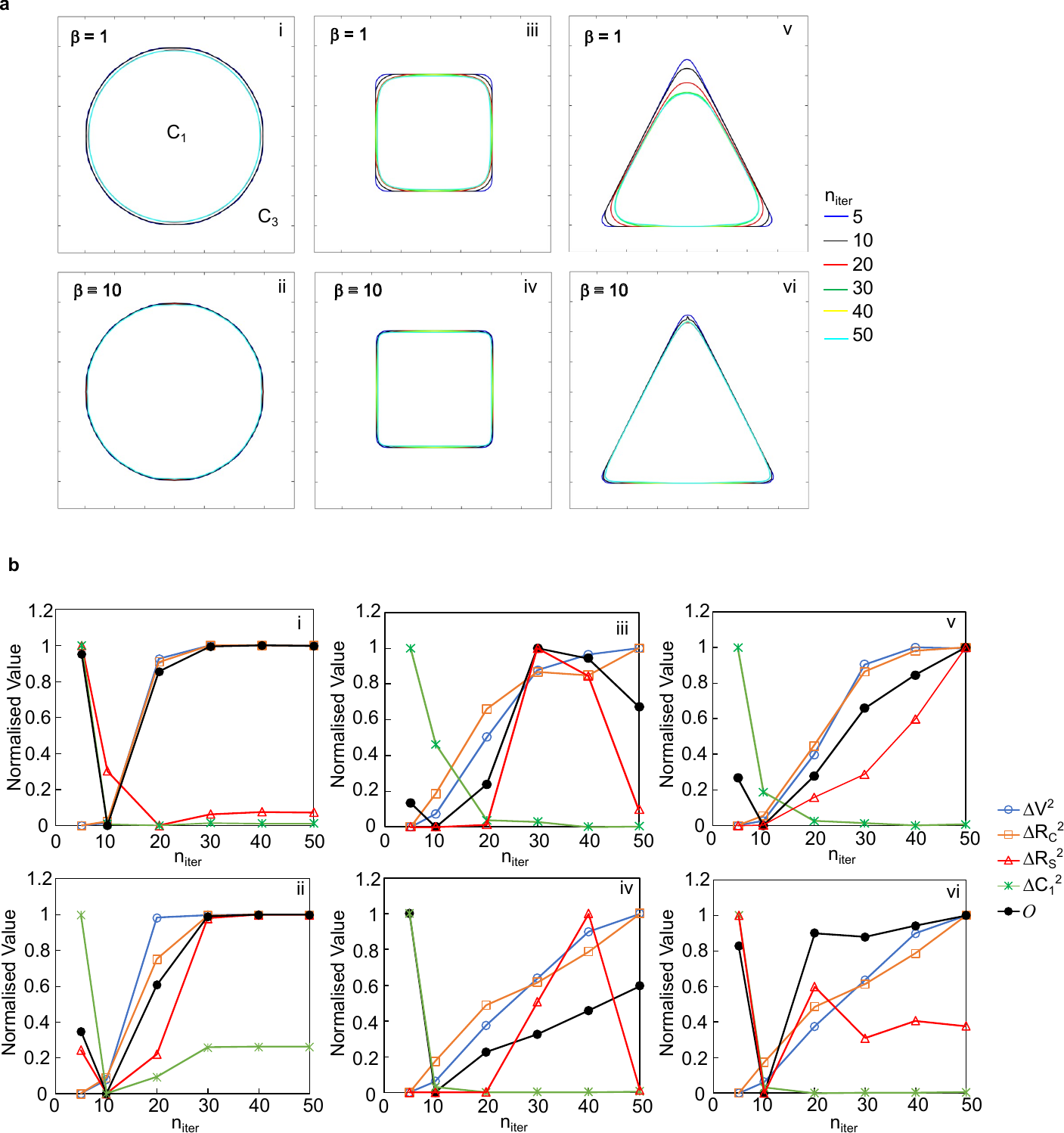}
	\caption{(a) 2D cross-section of the solid-gas interface of sphere (panels i and ii), cube (panels iii and iv) and cone (panels v and vi) geometries after different number of iterations in the first stage of energy minimisation with $\beta = 1$ (top row) and $\beta = 10$ (bottom row). (b) Plots of normalised values of the objection function's variables against the number of iterations for the corresponding cases in panel (a). }
	\label{fig:parametric_analysis}
\end{figure}

To determine the effective value of $\beta$ and the number of iterations, $n_{\textrm{iter}}$, required to form a desired solid geometry in the first stage of the energy minimisation process, we perform a survey based on several criteria: the size of the solid ($\Delta V$) referred to as the difference in the volume (area) between the simulated and expected solid shape in 3D (2D), the closeness of the shape to the intended one ($\Delta R_C$) referred to as the difference between the radius of the simulated and expected points at the interface of the solid, the smoothness of the interface ($\Delta R_S$) referred to as the difference between the radius of the simulated and fitted points at the solid interface, and the fitness of the tanh profile of the solid component across a gas-solid interface to the analytical solution ($\Delta C_1$). These criteria are chosen to ensure the resulting solid geometry has a smooth interface and maintains the intended shape and size. These criteria are then translated into an objective function that measures the square of the relative error of each criterion, given by
\begin{equation}
O = (\Delta V)^2 + (\Delta R_C)^2 + (\Delta R_S)^2 + (\Delta C_1)^2 .   
\label{frozen:scoring-function}
\end{equation}
All variables are then normalised to 1 to ensure that each criterion is equally favourable. This is done by normalising the root mean square error (RMSE) of each variable using the formula: $(RMSE_i - \min(RMSE_i) )/ (\max(RMSE_i)-\min(RMSE_i))$, where $i$ denotes for each variable in the function $O$. This objective function is then minimised against several solid geometries including sphere, cube and cone shapes to account for the geometry's curved, flat, inclined and corner features as shown in Fig.~\ref{fig:parametric_analysis}.

Here, we show two values of $\beta$ to represent weak ($\beta = 1$) and strong ($\beta = 10$) confining potentials. The energy minimisation steps are done up to $n_{\textrm{iter}}$ = 50. The snapshots of the 2D cross-section of the solid-gas interface for each shape for $n_{\textrm{iter}} = 5, 10, 20, 30, 40$ and $50$ are shown in Fig.~\ref{fig:parametric_analysis}(a). From these images, the impact of the value of $\beta$ can be seen primarily in maintaining the size and shape of the geometry. For $\beta = 1$, the size of the geometry shrinks as the iteration increases. This is prominent for shapes with corner features as in the cube and cone geometries. This indicates that two adjection solid interfaces separated by a narrow gap might merge into one solid interface if weak confining potential is used. A stronger confining potential ($\beta = 10$) is seen to be able to maintain the size and shape of the geometry as the number of iterations increases. However, it can lead to reducing the interface's smoothness and the interface's tanh profile fitness if the number of iterations is large. The effect of the number of interations can be evaluated further from the objective function analysis, as shown in Fig.\ref{fig:parametric_analysis}(b). Here, all variables in the objective function have been normalised such that the values range from 0 to 1 to ensure that each variable contributes similarly to the overall objective function. Combining all variables into the total objective function, $O$, we can see that the minimum deviation is obtained at 10 iterations for both values of $\beta$, which indicates the optimum solid geometry.

\begin{figure}[h!]
	\centering
	\includegraphics[width=16cm]{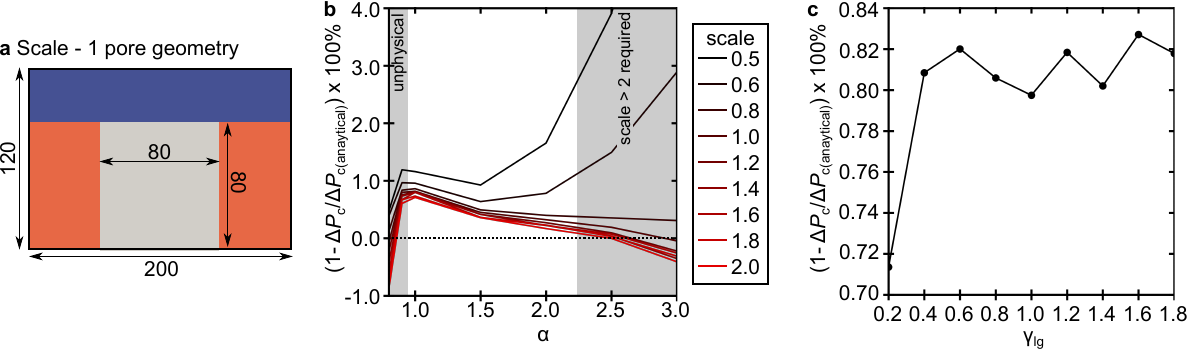}
	\caption{(a) Default (scale - 1) geometry of the 2D rectangular pore used to examine the impact of $\alpha$, domain scale, and $\gamma_{\mathrm{lg}}$ on the critical intrusion pressure.
    (b) The impact of $\alpha$ and the domain scale on the percentage difference between the simulated critical pressure, $\Delta P_{\mathrm{c}}$, and analytical critical pressure, $\Delta P_{\mathrm{c(analytical)}}$. The grey 'suitable zone' region highlighted indicates the range of $\alpha$ capable of producing physically sensible liquid-gas interfaces within the range of domain scales tested. (c) The deviation in the critical pressure from the analytical value for the scale-1.0, $\alpha=1.0$ system at a range of different liquid-gas surface tensions.}
	\label{fig:alpha_scale_gamma}
\end{figure}

\subsubsection{Fluid and domain parameters}

To assess the impact of the interfacial properties on the accuracy of simulated wetting characteristics, the critical intrusion pressure is examined for a 2D rectangular pore, shown in Fig.~\ref{fig:alpha_scale_gamma}. The parameters tested are the interfacial width, $\alpha$, domain scale, and liquid-gas interfacial tension, $\gamma_{\mathrm{lg}}$. The contact angle remains fixed throughout at $\theta=120^{\circ}$, and the grid spacing remains fixed at 1.0. In Fig.~\ref{fig:alpha_scale_gamma}(a), the default (scale-1) system dimensions are established (all measurements reported in lattice units). To examine the influence of system scale, all dimensions in this default domain are scaled by a chosen factor.  

Using an identical procedure to that described in the main text (Section: Critical Pressure on Micropillared Surfaces), we find the critical intrusion pressure required for liquid to spontaneously fill the pore, $\Delta P_{\mathrm{c}}$. This is then compared to the analytical critical intrusion pressure, $\Delta P_{\mathrm{c(analytical)}}$, defined in an identical manner to main text Eq. (34):
\begin{equation}
\Delta P_{\mathrm{c(analytical)}} = -\frac{2\gamma_{\textrm{lg}}\cos{\theta}}{s},
\end{equation}
where $s$ is the pore width. In Fig.~\ref{fig:alpha_scale_gamma}(b), this comparison between simulated and analytical critical pressures is made across a range of $\alpha$ and system scales ($\gamma_{\mathrm{lg}}$ is fixed at 1.0). A successful simulation must balance accuracy with computational expense. For $\alpha<0.9$ the interfacial profile is under-resolved, leading to unphysical interface curvatures. For $\alpha \geq 0.9$ (where the interfacial profile is sufficiently resolved), a compromise exists: to achieve highly accurate simulations, a larger $\alpha$ is required, but this in turn requires a larger system scale. The latter is required as for the mesoscale wetting phenomena investigated here, the interface width must remain small relative to the pore width. For highly accurate critical pressures, $\alpha=2.0$ is recommended with at least double the default domain scale. Greater accuracy can be achieved for $\alpha>2.0$, but these require larger system scales than presented in Fig.~\ref{fig:alpha_scale_gamma}(b). However, even for $\alpha=1.0$ at the default scale (1.0), the discrepancy between the simulated and analytical critical pressure is small (less than $1\%$). Given the computational efficiency associated with a smaller domain size, $\alpha=1.0$ (the same as the grid spacing) is judged to appropriately balance efficiency and accuracy, and is used throughout this work.

In Fig.~\ref{fig:alpha_scale_gamma}(c), the default system is used, but this time a range of $\gamma_{\mathrm{lg}}$ is tested. Across this range it can be seen that the accuracy remains close to 0.8$\%$.

\newpage

\subsection{Contact Angle Derivation}
Here we will describe the derivation of the surface tensions and equilibrium contact angle from the free energy given by 
\begin{equation}
F=\int\left[\frac{\kappa_2}{2}C_2^2(1-C_2)^2+\frac{\kappa_3}{2}C_3^2(1-C_3)^2 + \alpha^2\frac{\kappa_2}{2}(\boldsymbol{\nabla} C_2)^2+\alpha^2\frac{\kappa_3}{2}(\boldsymbol{\nabla} C_3)\right]\textrm{d}V.
\label{eqn:febinary}
\end{equation}
First we consider a pure interface between components $C_2$ and $C_3$. In this limit, $C_1=0$ and $\boldsymbol{\nabla} C_1=0$, so we arrive at a simplified free energy
\begin{equation}
F=\int\left[\frac{\kappa_2+\kappa_3}{2}(1-C_2)^2(C_2)^2+\alpha^2\frac{\kappa_2+\kappa_3}{2}(\boldsymbol{\nabla} C_2)^2\right]\textrm{d}V.
\end{equation}
We can calculate the surface tension from the excess free energy per unit area~\cite{LBM}, assuming the interface profile follows $C_2=\frac{1}{2}+\frac{1}{2}\tanh\frac{x}{2\alpha}$,
\begin{equation}
\label{s12}
\gamma_{23}=\int^\infty_{-\infty}\left[\frac{\kappa_2+\kappa_3}{2}(1-C_2)^2(C_2)^2+\alpha^2\frac{\kappa_2+\kappa_3}{2}(\boldsymbol{\nabla} C_2)^2\right]\textrm{d}x\hiderel=\frac{\alpha(\kappa_2+\kappa_3)}{6}.
\end{equation}
We can then repeat this process for the solid-fluid interfaces. For the interface between $C_1$ and $C_2$, we must have that $C_3=1-C_1-C_2=0$, so $C_1=1-C_2$
\begin{equation}
\label{Free Energy1}
F=\int\left[\frac{\kappa_2}{2}(1-C_2)^2(C_2)^2+\alpha^2\frac{\kappa_2+\kappa_3}{2}(\boldsymbol{\nabla} C_2)^2\right]\textrm{d}V.
\end{equation}
Then we arrive at the surface tension
\begin{equation}
\label{Free Energy2}
\gamma_{12}=\int^\infty_{-\infty}\left[\frac{\kappa_2}{2}(1-C_2)^2(C_2)^2+\alpha^2\frac{\kappa_2}{2}(\boldsymbol{\nabla} C_2)^2\right]\textrm{d}x\hiderel=\frac{\alpha\kappa_{2}}{6}.
\end{equation}
Finally, for the interface between $C_1$ and $C_3$, we must have $C_2=0$ and $\boldsymbol{\nabla} C_2=0$
\begin{equation}
\label{Free Energy3}
F=\int{\left[\frac{\kappa_3}{2}(1-C_3)^2(C_3)^2+\alpha^2\frac{\kappa_3}{2}(\boldsymbol{\nabla} C_3)^2\right]}\textrm{d}V,
\end{equation}
from which we arrive at
\begin{equation}
\label{Free Energy4}
\gamma_{13}=\int^\infty_{-\infty}\left[\frac{\kappa_3}{2}(1-C_3)^2(C_3)^2+\alpha^2\frac{\kappa_3}{2}(\boldsymbol{\nabla} C_3)^2\right]\textrm{d}x\hiderel=\frac{\alpha\kappa_{3}}{6}.
\end{equation}

Subsequently, we can utilise Young's contact angle relation
\begin{equation}
\cos\theta=\frac{\gamma_{13}-\gamma_{12}}{\gamma_{23}}
\label{eq:CA}
\end{equation}
to calculate the equilibrium contact angle in terms of $\kappa_1$ and $\kappa_2$
\begin{equation}
\cos\theta=\frac{\kappa_3-\kappa_2}{\kappa_2+\kappa_3}.
\end{equation}

\newpage

\subsection{Pseudocode}
\begin{figure}[h!]
	\centering
	\includegraphics[width=15cm]{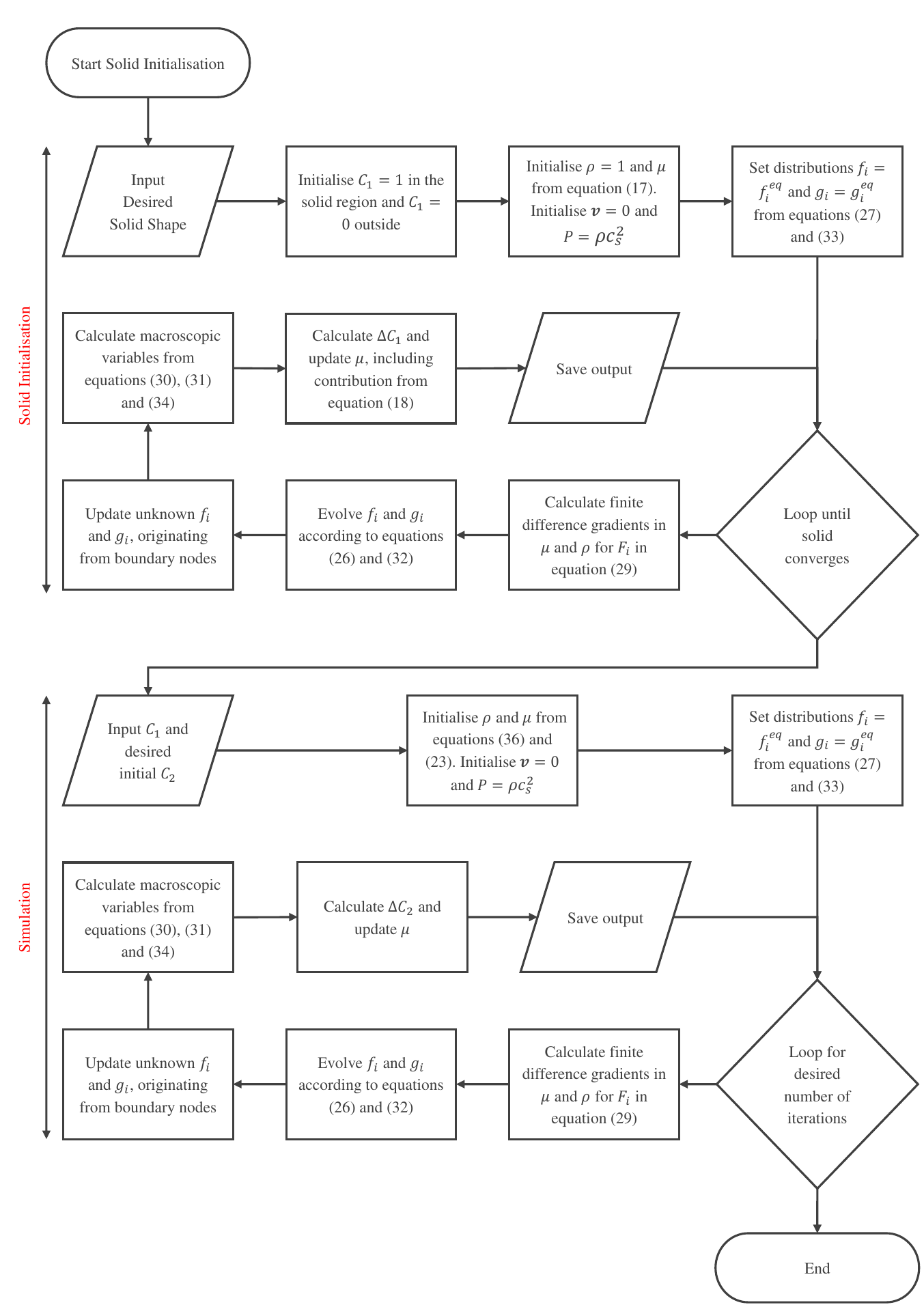}
	\label{fig:flowchary}
\end{figure}

\newpage

\subsection{Equilibrium Contact Angle Validation For Lattice Boltzmann Simulations}
\begin{figure}[h!]
	\centering
	\includegraphics[width=10cm]{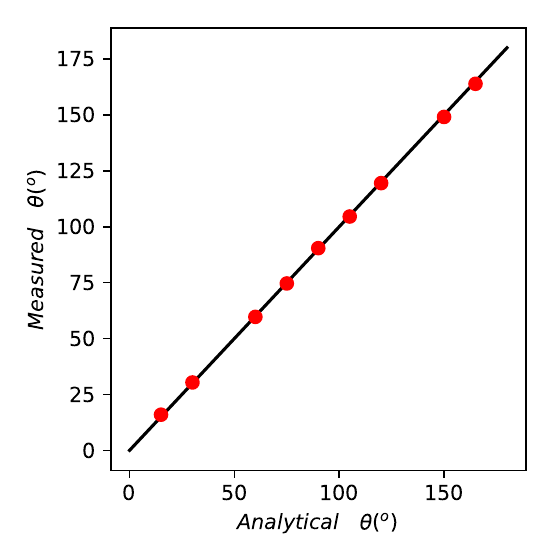}
	\caption{Measured vs theoretical contact angles in the lattice Boltzmann simulations for a droplet on a flat diffuse solid surface. The results agree to within $1.5^o$ for the range of contact angles tested ($15^o$ to $165^o$). The solid is allowed to equilibrate over 5000 time steps with $\alpha=1$ and $\beta=0.001$. A hemispherical droplet of initial radius $40$ is then placed on the surface and its equilibrium contact angle is changed by keeping $\gamma_{23}=\gamma_{12}=0.0025$ and modifying $\gamma_{13}$ in accordance with equation~(\ref{eq:CA}). As we are only interested in the accuracy of the final static result, we set $\tau_{2}=\tau_{3}=1.0$ and $\rho_{2}=\rho_{3}=1.0$.}
	\label{fig:capillary_filling_theta}
\end{figure}

\newpage

\subsection{\textit{Salvinia} sensitivity survey}

To evaluate the susceptibility of the critical intrusion pressure $\Delta P_{\mathrm{Int}}'$ and extrusion pressure $\Delta P_{\mathrm{Ext}}'$ to key parameters of the \textit{Salvinia} hair, a sensitivity analysis is performed. We start with the original model presented in the main text, featuring $D_{\mathrm{s}}/C_{\mathrm{s}}$ 0.77 (where $D_{\mathrm{s}}$ is the outermost hair diameter, and $C_{\mathrm{s}}$ is the centre-to-centre hair separation), a base contact angle $\theta_{\mathrm{base}}=160^\circ$, and a tip contact angle $\theta_{\mathrm{tip}}=35^\circ$. To explore the sensitivity of the critical pressures to these parameters, each parameter is varied by $\pm 5\%$, with the resulting critical pressures shown in Table I. Note that to vary the outermost diameter to centre-to-centre hair separation ratio, we maintain a fixed centre-to-centre hair separation, and radially scale the entire hair structure.
In Table I, it can be seen that the hair outer diameter has the largest influence on both critical pressures, compared to the contact angles. As anticipated from the analysis of the pinning locations in main text Fig. 5(f), the intruding interface is sensitive only to $\theta_{\mathrm{base}}$, whereas the extruding interface is sensitive only to $\theta_{\mathrm{tip}}$.

\begin{table}[h!]
\label{table:sensitivity}
\caption{Sensitivity analysis for the critical intrusion pressure $\Delta P_{\mathrm{Int}}'$ and extrusion pressure $\Delta P_{\mathrm{Ext}}'$ on the $Salvinia$ hair examined in the main text, where $D_{\mathrm{s}}$ is the outermost hair diameter, and $C_{\mathrm{s}}$ is the centre-to-centre hair separation.}
\begin{tabular}{c|cc|cc}
\hline
System & $\Delta P_{\mathrm{Int}}'$ & $\%$ increase & $\Delta P_{\mathrm{Ext}}'$ & $\%$ increase \\
\hline
Original & 2.94 && -0.82 & \\
\hline
$D_{\mathrm{s}}/C_{\mathrm{s}}$ $+5 \%$ & 3.26 & -10.97 & -0.83 & -1.72 \\
$D_{\mathrm{s}}/C_{\mathrm{s}}$ $-5 \%$ & 2.66 & 9.44 & -0.77 &	5.73 \\
\hline
$\theta_{\mathrm{base}}$ $+5 \%$ & 3.00 &	-2.04&	-0.82&	0.00\\
$\theta_{\mathrm{base}}$  $-5 \%$ & 2.87&	2.42&	-0.82&	0.00 \\
\hline
$\theta_{\mathrm{tip}}$  $+5 \%$ & 2.94&	0.00&	-0.81&	0.86 \\
$\theta_{\mathrm{tip}}$  $-5 \%$ & 2.94&	0.00&	-0.83&	-0.86\\
\hline
\end{tabular}
\end{table}

\newpage

\subsection{Dynamic Contact Angle in the Capillary Filling Benchmark}

\begin{figure}[h!]
	\centering
	\includegraphics[width=16cm]{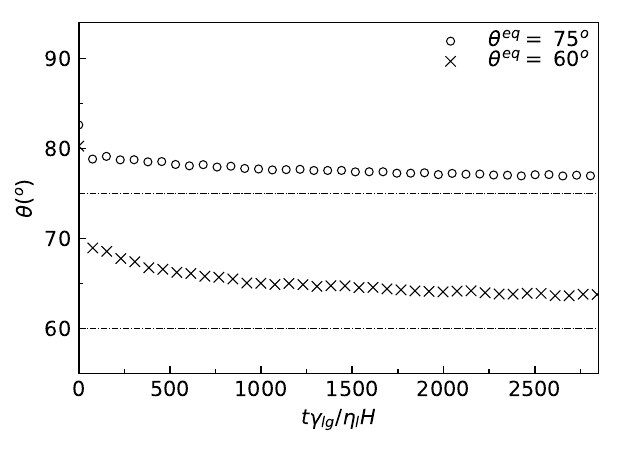}
	\caption{Dynamic contact angle of the liquid front in the capillary channel with time. The dynamic contact angle approaches the equilibrium value as the liquid fills the channel and the filling velocity decreases.}
	\label{fig:capillary_filling_theta_dynamic}
\end{figure}

\newpage

\bibliography{manuscript}

\end{document}